\documentclass[conference,10pt]{IEEEtran}
\usepackage[cmex10]{amsmath}
\usepackage[nolist,withpage]{acronym}
\usepackage{etoolbox}
\makeatletter
\newif\if@in@acrolist
\AtBeginEnvironment{acronym}{\@in@acrolisttrue}
\newrobustcmd{\LU}[2]{\if@in@acrolist#1\else#2\fi}

\newcommand{\ACF}[1]{{\@in@acrolisttrue\acf{#1}}}
\makeatother
\usepackage{amsthm,amsfonts,mathrsfs}
\usepackage{graphicx}
\usepackage{dsfont,enumitem,stackrel}
\usepackage{color}
\usepackage{mathtools,amssymb,bm,mathabx}
\usepackage{amstext}
\usepackage{array}
\usepackage{algorithmicx}
\usepackage[ruled]{algorithm}
\usepackage{algpseudocode}
\usepackage{algpascal}
\usepackage{algc}
\usepackage{subfigure}
\usepackage[font=footnotesize]{caption}
\usepackage{cases}
\usepackage{pgfplots}
\pgfplotsset{compat=newest}
\usetikzlibrary{spy}
\usepackage{accents}
\usepackage{float}
\usepackage{cite}

\theoremstyle{remark}

\newcommand\ASTART{\bigskip\noindent\begin{minipage}[b]{0.5\linewidth}}
	
	\newcommand\AENDSKIP{\end{minipage}\bigskip}
\newcommand\AEND{\end{minipage}}
\newcommand{\RN}[1]{%
	\textup{\uppercase\expandafter{\romannumeral#1}}%
}

\theoremstyle{plain}

\newtheorem{prop}{\textbf{Proposition}}
\newtheorem*{corl}{\textbf{Corollary}}

\theoremstyle{definition}

\theoremstyle{remark}

\newcommand*{\rom}[1]{\expandafter\@slowromancap\romannumeral #1@}

\definecolor{copperrose}{rgb}{0.6, 0.4, 0.4}
\definecolor{azure}{rgb}{0.0, 0.5, 1.0}
\definecolor{ashgrey}{rgb}{0.7, 0.75, 0.71}
\definecolor{chestnut}{rgb}{0.8, 0.36, 0.36}
\definecolor{airforceblue}{rgb}{0.36, 0.54, 0.66}
\definecolor{cadmiumorange}{rgb}{0.93, 0.53, 0.18}
\definecolor{bleudefrance}{rgb}{0.19, 0.55, 0.91}
\definecolor{carolinablue}{rgb}{0.6, 0.73, 0.89}
\definecolor{blue(ncs)}{rgb}{0.0, 0.53, 0.74}
\definecolor{dodgerblue}{rgb}{0.12, 0.56, 1.0}
\definecolor{cssgreen}{rgb}{0.0, 0.5, 0.0}
\definecolor{cadmiumgreen}{rgb}{0.0, 0.42, 0.24}
\definecolor{cadmiumorange}{rgb}{0.93, 0.53, 0.18}
\definecolor{amaranth}{rgb}{0.9, 0.17, 0.31}
\definecolor{bluegray}{rgb}{0.4, 0.6, 0.8}
\definecolor{cadmiumgreen}{rgb}{0.0, 0.42, 0.24}
\definecolor{cerulean}{rgb}{0.0, 0.48, 0.65}
\definecolor{ceil}{rgb}{0.57, 0.63, 0.81}

\title{Off-the-grid Blind Deconvolution and Demixing}
\author{\IEEEauthorblockN{Saeed Razavikia$^\dagger$, Sajad Daei$^\dagger$, Mikael Skoglund$^\dagger$, Gabor Fodor$^{\dagger,\flat}$, Carlo Fischione$^\dagger$}
 \IEEEauthorblockA{$^\dagger$School of Electrical Engineering and Computer Science, KTH Royal Institute of Technology, Stockholm, Sweden\\
 $^\flat$Ericsson Research, Sweden\\
Email: \{sraz, sajado, skoglund, gaborf, carlofi\}@kth.se}
\thanks{S. Razavikia was supported by the Wallenberg AI, Autonomous Systems and Software Program (WASP).}
}

\begin{document}
\begin{acronym}[LTE-Advanced]
\acro{GB2D}{gridless blind deconvolution and demixing}
  \acro{ISAC}{integrated (radar) sensing and communications}
  \acro{2G}{Second Generation}
  \acro{6G}{Sixth Generation}
  \acro{3-DAP}{3-Dimensional Assignment Problem}
  \acro{AA}{Antenna Array}
  \acro{AC}{Admission Control}
  \acro{AD}{Attack-Decay}
  \acro{ADC}{analog-to-digital conversion}
  \acro{ADMM}{alternating direction method of multipliers}
  \acro{ADSL}{Asymmetric Digital Subscriber Line}
  \acro{AHW}{Alternate Hop-and-Wait}
  \acro{AI}{Artificial Intelligence}
  \acro{AirComp}{Over-the-air computation}
  \acro{AMC}{Adaptive Modulation and Coding}
  \acro{AP}{\LU{A}{a}ccess \LU{P}{p}oint}
  \acro{APA}{Adaptive Power Allocation}
  \acro{ARMA}{Autoregressive Moving Average}
  \acro{ARQ}{\LU{A}{a}utomatic \LU{R}{r}epeat \LU{R}{r}equest}
  \acro{ATES}{Adaptive Throughput-based Efficiency-Satisfaction Trade-Off}
  \acro{AWGN}{additive white Gaussian noise}
  \acro{BAA}{\LU{B}{b}roadband \LU{A}{a}nalog \LU{A}{a}ggregation}
  \acro{BB}{Branch and Bound}
  \acro{BCD}{block coordinate descent}
  \acro{BD}{Block Diagonalization}
  \acro{BER}{Bit Error Rate}
  \acro{BF}{Best Fit}
  \acro{BFD}{bidirectional full duplex}
  \acro{BLER}{BLock Error Rate}
  \acro{BPC}{Binary Power Control}
  \acro{BPSK}{Binary Phase-Shift Keying}
  \acro{BRA}{Balanced Random Allocation}
  \acro{BS}{base station}
  \acro{BSUM}{block successive upper-bound minimization}
  \acro{CAP}{Combinatorial Allocation Problem}
  \acro{CAPEX}{Capital Expenditure}
  \acro{CBF}{Coordinated Beamforming}
  \acro{CBR}{Constant Bit Rate}
  \acro{CBS}{Class Based Scheduling}
  \acro{CC}{Congestion Control}
  \acro{CDF}{Cumulative Distribution Function}
  \acro{CDMA}{Code-Division Multiple Access}
  \acro{CE}{\LU{C}{c}hannel \LU{E}{e}stimation}
  \acro{CL}{Closed Loop}
  \acro{CLPC}{Closed Loop Power Control}
  \acro{CML}{centralized machine learning}
  \acro{CNR}{Channel-to-Noise Ratio}
  \acro{CNN}{\LU{C}{c}onvolutional \LU{N}{n}eural \LU{N}{n}etwork}
  \acro{CPA}{Cellular Protection Algorithm}
  \acro{CPICH}{Common Pilot Channel}
  \acro{CoCoA}{\LU{C}{c}ommunication efficient distributed dual \LU{C}{c}oordinate \LU{A}{a}scent}
  \acro{CoMAC}{\LU{C}{c}omputation over \LU{M}{m}ultiple-\LU{A}{a}ccess \LU{C}{c}hannels}
  \acro{CoMP}{Coordinated Multi-Point}
  \acro{CQI}{Channel Quality Indicator}
  \acro{CRM}{Constrained Rate Maximization}
	\acro{CRN}{Cognitive Radio Network}
  \acro{CS}{Coordinated Scheduling}
  \acro{CSI}{\LU{C}{c}hannel \LU{S}{s}tate \LU{I}{i}nformation}
  \acro{CSMA}{\LU{C}{c}arrier \LU{S}{s}ense \LU{M}{m}ultiple \LU{A}{a}ccess}
  \acro{CUE}{Cellular User Equipment}
  \acro{D2D}{device-to-device}
  \acro{DAC}{digital-to-analog converter}
  \acro{DC}{direct current}
  \acro{DCA}{Dynamic Channel Allocation}
  \acro{DE}{Differential Evolution}
  \acro{DFT}{Discrete Fourier Transform}
  \acro{DIST}{Distance}
  \acro{DL}{downlink}
  \acro{DMA}{Double Moving Average}
  \acro{DML}{Distributed Machine Learning}
  \acro{DMRS}{demodulation reference signal}
  \acro{D2DM}{D2D Mode}
  \acro{DMS}{D2D Mode Selection}
  \acro{DNN}{Deep Neural Network}
  \acro{DPC}{Dirty Paper Coding}
  \acro{DRA}{Dynamic Resource Assignment}
  \acro{DSA}{Dynamic Spectrum Access}
  \acro{DSGD}{\LU{D}{d}istributed \LU{S}{s}tochastic \LU{G}{g}radient \LU{D}{d}escent}
  \acro{DSM}{Delay-based Satisfaction Maximization}
  \acro{ECC}{Electronic Communications Committee}
  \acro{EFLC}{Error Feedback Based Load Control}
  \acro{EI}{Efficiency Indicator}
  \acro{eNB}{Evolved Node B}
  \acro{EPA}{Equal Power Allocation}
  \acro{EPC}{Evolved Packet Core}
  \acro{EPS}{Evolved Packet System}
  \acro{E-UTRAN}{Evolved Universal Terrestrial Radio Access Network}
  \acro{ES}{Exhaustive Search}
  \acro{FC}{\LU{F}{f}usion \LU{C}{c}enter}
  \acro{FD}{\LU{F}{f}ederated \LU{D}{d}istillation}
  \acro{FDD}{frequency division duplex}
  \acro{FDM}{Frequency Division Multiplexing}
  \acro{FDMA}{\LU{F}{f}requency \LU{D}{d}ivision \LU{M}{m}ultiple \LU{A}{a}ccess}
  \acro{FedAvg}{\LU{F}{f}ederated \LU{A}{a}veraging}
  \acro{FER}{Frame Erasure Rate}
  \acro{FF}{Fast Fading}
  \acro{FL}{Federated Learning}
  \acro{FSB}{Fixed Switched Beamforming}
  \acro{FST}{Fixed SNR Target}
  \acro{FTP}{File Transfer Protocol}
  \acro{GA}{Genetic Algorithm}
  \acro{GBR}{Guaranteed Bit Rate}
  \acro{GD}{gradient descent}
  \acro{GLR}{Gain to Leakage Ratio}
  \acro{GOS}{Generated Orthogonal Sequence}
  \acro{GPL}{GNU General Public License}
  \acro{GRP}{Grouping}
  \acro{HARQ}{Hybrid Automatic Repeat Request}
  \acro{HD}{half-duplex}
  \acro{HMS}{Harmonic Mode Selection}
  \acro{HOL}{Head Of Line}
  \acro{HSDPA}{High-Speed Downlink Packet Access}
  \acro{HSPA}{High Speed Packet Access}
  \acro{HTTP}{HyperText Transfer Protocol}
  \acro{ICMP}{Internet Control Message Protocol}
  \acro{ICI}{Intercell Interference}
  \acro{ID}{Identification}
  \acro{IETF}{Internet Engineering Task Force}
  \acro{ILP}{Integer Linear Program}
  \acro{JRAPAP}{Joint RB Assignment and Power Allocation Problem}
  \acro{UID}{Unique Identification}
  \acro{IID}{\LU{I}{i}ndependent and \LU{I}{i}dentically \LU{D}{d}istributed}
  \acro{IIR}{Infinite Impulse Response}
  \acro{ILP}{Integer Linear Problem}
  \acro{IMT}{International Mobile Telecommunications}
  \acro{INV}{Inverted Norm-based Grouping}
  \acro{IoT}{Internet of Things}
  \acro{IP}{Integer Programming}
  \acro{IPv6}{Internet Protocol Version 6}
  \acro{IQ}{in-phase quadrature}
  \acro{ISD}{Inter-Site Distance}
  \acro{ISI}{Inter Symbol Interference}
  \acro{ITU}{International Telecommunication Union}
  \acro{JAFM}{joint assignment and fairness maximization}
  \acro{JAFMA}{joint assignment and fairness maximization algorithm}
  \acro{JOAS}{Joint Opportunistic Assignment and Scheduling}
  \acro{JOS}{Joint Opportunistic Scheduling}
  \acro{JP}{Joint Processing}
	\acro{JS}{Jump-Stay}
  \acro{KKT}{Karush-Kuhn-Tucker}
  \acro{L3}{Layer-3}
  \acro{LAC}{Link Admission Control}
  \acro{LA}{Link Adaptation}
  \acro{LC}{Load Control}
  \acro{LDC}{\LU{L}{l}earning-\LU{D}{d}riven \LU{C}{c}ommunication}
  \acro{LOS}{line of sight}
  \acro{LP}{Linear Programming}
  \acro{LTE}{Long Term Evolution}
	\acro{LTE-A}{\ac{LTE}-Advanced}
  \acro{LTE-Advanced}{Long Term Evolution Advanced}
  \acro{M2M}{Machine-to-Machine}
  \acro{MAC}{multiple access channel}
  \acro{MANET}{Mobile Ad hoc Network}
  \acro{MC}{Modular Clock}
  \acro{MCS}{Modulation and Coding Scheme}
  \acro{MDB}{Measured Delay Based}
  \acro{MDI}{Minimum D2D Interference}
  \acro{MF}{Matched Filter}
  \acro{MG}{Maximum Gain}
  \acro{MH}{Multi-Hop}
  \acro{MIMO}{\LU{M}{m}ultiple \LU{I}{i}nput \LU{M}{m}ultiple \LU{O}{o}utput}
  \acro{MINLP}{mixed integer nonlinear programming}
  \acro{MIP}{Mixed Integer Programming}
  \acro{MISO}{multiple input single output}
  \acro{ML}{Machine Learning}
  \acro{MLWDF}{Modified Largest Weighted Delay First}
  \acro{MME}{Mobility Management Entity}
  \acro{MMSE}{minimum mean squared error}
  \acro{MOS}{Mean Opinion Score}
  \acro{MPF}{Multicarrier Proportional Fair}
  \acro{MRA}{Maximum Rate Allocation}
  \acro{MR}{Maximum Rate}
  \acro{MRC}{Maximum Ratio Combining}
  \acro{MRT}{Maximum Ratio Transmission}
  \acro{MRUS}{Maximum Rate with User Satisfaction}
  \acro{MS}{Mode Selection}
  \acro{MSE}{\LU{M}{m}ean \LU{S}{s}quared \LU{E}{e}rror}
  \acro{MSI}{Multi-Stream Interference}
  \acro{MTC}{Machine-Type Communication}
  \acro{MTSI}{Multimedia Telephony Services over IMS}
  \acro{MTSM}{Modified Throughput-based Satisfaction Maximization}
  \acro{MU-MIMO}{Multi-User Multiple Input Multiple Output}
  \acro{MU}{Multi-User}
  \acro{NAS}{Non-Access Stratum}
  \acro{NB}{Node B}
	\acro{NCL}{Neighbor Cell List}
  \acro{NLP}{Nonlinear Programming}
  \acro{NLOS}{non-line of sight}
  \acro{NMSE}{Normalized Mean Square Error}
  \acro{NN}{Neural Network}
  \acro{NOMA}{\LU{N}{n}on-\LU{O}{o}rthogonal \LU{M}{m}ultiple \LU{A}{a}ccess}
  \acro{NORM}{Normalized Projection-based Grouping}
  \acro{NP}{non-polynomial time}
  \acro{NRT}{Non-Real Time}
  \acro{NSPS}{National Security and Public Safety Services}
  \acro{O2I}{Outdoor to Indoor}
  \acro{OAC}{Over-the-Air Computation}
  \acro{OFDMA}{\LU{O}{o}rthogonal \LU{F}{f}requency \LU{D}{d}ivision \LU{M}{m}ultiple \LU{A}{a}ccess}
  \acro{OFDM}{Orthogonal Frequency Division Multiplexing}
  \acro{OFPC}{Open Loop with Fractional Path Loss Compensation}
	\acro{O2I}{Outdoor-to-Indoor}
  \acro{OL}{Open Loop}
  \acro{OLPC}{Open-Loop Power Control}
  \acro{OL-PC}{Open-Loop Power Control}
  \acro{OPEX}{Operational Expenditure}
  \acro{ORB}{Orthogonal Random Beamforming}
  \acro{JO-PF}{Joint Opportunistic Proportional Fair}
  \acro{OSI}{Open Systems Interconnection}
  \acro{PAIR}{D2D Pair Gain-based Grouping}
  \acro{PAPR}{Peak-to-Average Power Ratio}
  \acro{P2P}{Peer-to-Peer}
  \acro{PC}{Power Control}
  \acro{PCI}{Physical Cell ID}
  \acro{PDCCH}{physical downlink control channel}
  \acro{PDD}{penalty dual decomposition}
  \acro{PDF}{Probability Density Function}
  \acro{PER}{Packet Error Rate}
  \acro{PF}{Proportional Fair}
  \acro{P-GW}{Packet Data Network Gateway}
  \acro{PL}{Pathloss}
  \acro{PLL}{phase-locked Loop}
  \acro{PRB}{Physical Resource Block}
  \acro{PROJ}{Projection-based Grouping}
  \acro{ProSe}{Proximity Services}
  \acro{PS}{\LU{P}{p}arameter \LU{S}{s}erver}
  \acro{PSO}{Particle Swarm Optimization}
  \acro{PUCCH}{physical uplink control channel}
  \acro{PZF}{Projected Zero-Forcing}
  \acro{QAM}{Quadrature Amplitude Modulation}
  \acro{QoS}{quality of service}
  \acro{QPSK}{Quadri-Phase Shift Keying}
  \acro{RAISES}{Reallocation-based Assignment for Improved Spectral Efficiency and Satisfaction}
  \acro{RAN}{Radio Access Network}
  \acro{RA}{Resource Allocation}
  \acro{RAT}{Radio Access Technology}
  \acro{RATE}{Rate-based}
  \acro{RB}{resource block}
  \acro{RBG}{Resource Block Group}
  \acro{REF}{Reference Grouping}
  \acro{RF}{radio frequency}
  \acro{RLC}{Radio Link Control}
  \acro{RM}{Rate Maximization}
  \acro{RNC}{Radio Network Controller}
  \acro{RND}{Random Grouping}
  \acro{RRA}{Radio Resource Allocation}
  \acro{RRM}{\LU{R}{r}adio \LU{R}{r}esource \LU{M}{m}anagement}
  \acro{RSCP}{Received Signal Code Power}
  \acro{RSRP}{reference signal receive power}
  \acro{RSRQ}{Reference Signal Receive Quality}
  \acro{RR}{Round Robin}
  \acro{RRC}{Radio Resource Control}
  \acro{RSSI}{received signal strength indicator}
  \acro{RT}{Real Time}
  \acro{RU}{Resource Unit}
  \acro{RUNE}{RUdimentary Network Emulator}
  \acro{RV}{Random Variable}
  \acro{SAC}{Session Admission Control}
  \acro{SCM}{Spatial Channel Model}
  \acro{SC-FDMA}{Single Carrier - Frequency Division Multiple Access}
  \acro{SD}{Soft Dropping}
  \acro{S-D}{Source-Destination}
  \acro{SDPC}{Soft Dropping Power Control}
  \acro{SDMA}{Space-Division Multiple Access}
  \acro{SDR}{semidefinite relaxation}
  \acro{SDP}{semidefinite programming}
  \acro{SER}{Symbol Error Rate}
  \acro{SES}{Simple Exponential Smoothing}
  \acro{S-GW}{Serving Gateway}
  \acro{SGD}{\LU{S}{s}tochastic \LU{G}{g}radient \LU{D}{d}escent}  
  \acro{SINR}{signal-to-interference-plus-noise ratio}
  \acro{SI}{self-interference}
  \acro{SIP}{Session Initiation Protocol}
  \acro{SISO}{\LU{S}{s}ingle \LU{I}{i}nput \LU{S}{s}ingle \LU{O}{o}utput}
  \acro{SIMO}{Single Input Multiple Output}
  \acro{SIR}{Signal to Interference Ratio}
  \acro{SLNR}{Signal-to-Leakage-plus-Noise Ratio}
  \acro{SMA}{Simple Moving Average}
  \acro{SNR}{\LU{S}{s}ignal-to-\LU{N}{n}oise \LU{R}{r}atio}
  \acro{SORA}{Satisfaction Oriented Resource Allocation}
  \acro{SORA-NRT}{Satisfaction-Oriented Resource Allocation for Non-Real Time Services}
  \acro{SORA-RT}{Satisfaction-Oriented Resource Allocation for Real Time Services}
  \acro{SPF}{Single-Carrier Proportional Fair}
  \acro{SRA}{Sequential Removal Algorithm}
  \acro{SRS}{sounding reference signal}
  \acro{SU-MIMO}{Single-User Multiple Input Multiple Output}
  \acro{SU}{Single-User}
  \acro{SVD}{Singular Value Decomposition}
  \acro{SVM}{\LU{S}{s}upport \LU{V}{v}ector \LU{M}{m}achine}
  \acro{TCP}{Transmission Control Protocol}
  \acro{TDD}{time division duplex}
  \acro{TDMA}{\LU{T}{t}ime \LU{D}{d}ivision \LU{M}{m}ultiple \LU{A}{a}ccess}
  \acro{TNFD}{three node full duplex}
  \acro{TETRA}{Terrestrial Trunked Radio}
  \acro{TP}{Transmit Power}
  \acro{TPC}{Transmit Power Control}
  \acro{TTI}{transmission time interval}
  \acro{TTR}{Time-To-Rendezvous}
  \acro{TSM}{Throughput-based Satisfaction Maximization}
  \acro{TU}{Typical Urban}
  \acro{UE}{\LU{U}{u}ser \LU{E}{e}quipment}
  \acro{UEPS}{Urgency and Efficiency-based Packet Scheduling}
  \acro{UL}{uplink}
  \acro{UMTS}{Universal Mobile Telecommunications System}
  \acro{URI}{Uniform Resource Identifier}
  \acro{URM}{Unconstrained Rate Maximization}
  \acro{VR}{Virtual Resource}
  \acro{VoIP}{Voice over IP}
  \acro{WAN}{Wireless Access Network}
  \acro{WCDMA}{Wideband Code Division Multiple Access}
  \acro{WF}{Water-filling}
  \acro{WiMAX}{Worldwide Interoperability for Microwave Access}
  \acro{WINNER}{Wireless World Initiative New Radio}
  \acro{WLAN}{Wireless Local Area Network}
  \acro{WMMSE}{weighted minimum mean square error}
  \acro{WMPF}{Weighted Multicarrier Proportional Fair}
  \acro{WPF}{Weighted Proportional Fair}
  \acro{WSN}{Wireless Sensor Network}
  \acro{WWW}{World Wide Web}
  \acro{XIXO}{(Single or Multiple) Input (Single or Multiple) Output}
  \acro{ZF}{Zero-Forcing}
  \acro{ZMCSCG}{Zero Mean Circularly Symmetric Complex Gaussian}
\end{acronym}

\bstctlcite{IEEEexample:BSTcontrol}
\maketitle

\begin{abstract}
We consider the problem of \ac{GB2D} in scenarios where multiple users communicate messages through multiple unknown channels, and a single \ac{BS} collects their contributions. This scenario arises in various communication fields, including wireless communications, the Internet of Things, over-the-air computation, and integrated sensing and communications. In this setup, each user's message is convolved with a multi-path channel formed by several scaled and delayed copies of Dirac spikes. The \ac{BS} receives a linear combination of the convolved signals, and the goal is to recover the unknown amplitudes, continuous-indexed delays, and transmitted waveforms from a compressed vector of measurements at the \ac{BS}. However, in the absence of any prior knowledge of the transmitted messages and channels, \ac{GB2D} is highly challenging and intractable in general. To address this issue, we assume that each user's message follows a distinct modulation scheme living in a known low-dimensional subspace. By exploiting these subspace assumptions and the sparsity of the multipath channels for different users, we transform the nonlinear \ac{GB2D} problem into a matrix tuple recovery problem from a few linear measurements. To achieve this, we propose a semidefinite programming optimization that exploits the specific low-dimensional structure of the matrix tuple to recover the messages and continuous delays of different communication paths from a single received signal at the \ac{BS}. Finally, our numerical experiments show that our proposed method effectively recovers all transmitted messages and the continuous delay parameters of the channels with a sufficient number of samples.
\end{abstract}


\makeatletter{\renewcommand*{\@makefnmark}{}
\footnotetext{Saeed Razavikia and Carlo Fischione acknowledge the support of WASP, SSF SAICOM, VR, EU FLASH project, and Digital Futures.

Sajad Daei and Mikael Skoglund were supported in part by Digital Futures. Digital Futures Project PERCy supported the work of Gabor Fodor. 

 }
\makeatother}

\section{Introduction}
In the near future, the Internet of Things (IoT) is expected to connect billions of wireless devices, surpassing the capacity of the current fifth-generation (5G) wireless system both technically and economically. One of the primary challenges that 6G, the future wireless communication system, will face is managing the massive number of IoT devices that generate sporadic traffic. As the 6G market grows, this sporadic traffic will significantly increase, and it is generally agreed among communications engineers that the current 5G channel access procedures cannot handle this volume of traffic.

Traditional channel access methods, which rely on classical information and communication theory, require a large number of pilots or training signals to estimate the channel, leading to significant resource waste that does not scale towards IoT requirements. Thus, minimizing the overhead caused by exchanging certain types of training information, such as channel estimation and data slot assignment, is necessary. This is especially critical for communications over dynamic channels, such as millimeter-wave or terahertz, where channel coherence times are short, and the channel state information changes rapidly. In these cases, the assumption of block fading no longer holds. One approach to addressing this issue is to incorporate channel aging effects into the channel estimation process to maximize spectral efficiency (see, e.g., \cite{fodor2023optimizing}), but this requires knowledge of the channel correlation structure at different times, which might be challenging to obtain in general channel environments. Therefore, for situations where a large number of devices transmit small amounts of data sporadically over dynamic channels, and the channel correlation structure is unknown, it is crucial to avoid transmitting a signal with much longer overhead information than actual data. This raises the question of whether this is feasible.

 \begin{figure*}
    \centering

\scalebox{1}{

\tikzset{every picture/.style={line width=0.75pt}} 

\begin{tikzpicture}[x=0.75pt,y=0.75pt,yscale=-1,xscale=1]

\draw    (262.5,45) -- (297.14,72.15) ;
\draw [shift={(299.5,74)}, rotate = 218.09] [fill={rgb, 255:red, 0; green, 0; blue, 0 }  ][line width=0.08]  [draw opacity=0] (5.36,-2.57) -- (0,0) -- (5.36,2.57) -- (3.56,0) -- cycle    ;
\draw    (261.5,117) -- (296.2,87.93) ;
\draw [shift={(298.5,86)}, rotate = 140.04] [fill={rgb, 255:red, 0; green, 0; blue, 0 }  ][line width=0.08]  [draw opacity=0] (5.36,-2.57) -- (0,0) -- (5.36,2.57) -- (3.56,0) -- cycle    ;
\draw   (296,79.5) .. controls (296,75.36) and (299.25,72) .. (303.25,72) .. controls (307.25,72) and (310.5,75.36) .. (310.5,79.5) .. controls (310.5,83.64) and (307.25,87) .. (303.25,87) .. controls (299.25,87) and (296,83.64) .. (296,79.5) -- cycle ; \draw   (296,79.5) -- (310.5,79.5) ; \draw   (303.25,72) -- (303.25,87) ;
\draw    (311.5,79) -- (329.5,79) ;
\draw [shift={(332.5,79)}, rotate = 180] [fill={rgb, 255:red, 0; green, 0; blue, 0 }  ][line width=0.08]  [draw opacity=0] (5.36,-2.57) -- (0,0) -- (5.36,2.57) -- (3.56,0) -- cycle    ;
\draw  [color={rgb, 255:red, 74; green, 74; blue, 74 }  ,draw opacity=1 ][fill={rgb, 1:red, 0.7; green, 0.75; blue, 0.71 }  ,fill opacity=0.62 ] (529,68.6) .. controls (529,64.4) and (532.4,61) .. (536.6,61) -- (567.9,61) .. controls (572.1,61) and (575.5,64.4) .. (575.5,68.6) -- (575.5,91.4) .. controls (575.5,95.6) and (572.1,99) .. (567.9,99) -- (536.6,99) .. controls (532.4,99) and (529,95.6) .. (529,91.4) -- cycle ;
\draw    (576.5,81) -- (600.5,80.65) ;
\draw [shift={(603.5,80.61)}, rotate = 179.17] [fill={rgb, 255:red, 0; green, 0; blue, 0 }  ][line width=0.08]  [draw opacity=0] (5.36,-2.57) -- (0,0) -- (5.36,2.57) -- (3.56,0) -- cycle    ;
\draw    (501.5,81) -- (525.5,80.65) ;
\draw [shift={(528.5,80.61)}, rotate = 179.17] [fill={rgb, 255:red, 0; green, 0; blue, 0 }  ][line width=0.08]  [draw opacity=0] (5.36,-2.57) -- (0,0) -- (5.36,2.57) -- (3.56,0) -- cycle    ;
\draw    (53.5,124) -- (71.5,124) ;
\draw [shift={(74.5,124)}, rotate = 180] [fill={rgb, 255:red, 0; green, 0; blue, 0 }  ][line width=0.08]  [draw opacity=0] (5.36,-2.57) -- (0,0) -- (5.36,2.57) -- (3.56,0) -- cycle    ;
\draw    (55.5,35) -- (73.5,35) ;
\draw [shift={(76.5,35)}, rotate = 180] [fill={rgb, 255:red, 0; green, 0; blue, 0 }  ][line width=0.08]  [draw opacity=0] (5.36,-2.57) -- (0,0) -- (5.36,2.57) -- (3.56,0) -- cycle    ;
\draw    (125.5,35) -- (143.5,35) ;
\draw [shift={(146.5,35)}, rotate = 180] [fill={rgb, 255:red, 0; green, 0; blue, 0 }  ][line width=0.08]  [draw opacity=0] (5.36,-2.57) -- (0,0) -- (5.36,2.57) -- (3.56,0) -- cycle    ;
\draw    (122.5,124) -- (140.5,124) ;
\draw [shift={(143.5,124)}, rotate = 180] [fill={rgb, 255:red, 0; green, 0; blue, 0 }  ][line width=0.08]  [draw opacity=0] (5.36,-2.57) -- (0,0) -- (5.36,2.57) -- (3.56,0) -- cycle    ;
\draw    (322.5,135) -- (373.5,135) ;
\draw [line width=0.75]    (353.35,134.98) -- (353.35,112.52) ;
\draw [shift={(353.35,109.52)}, rotate = 90] [fill={rgb, 255:red, 0; green, 0; blue, 0 }  ][line width=0.08]  [draw opacity=0] (5.36,-2.57) -- (0,0) -- (5.36,2.57) -- cycle    ;
\draw [line width=0.75]    (331.12,135) -- (331.01,108) ;
\draw [shift={(331,105)}, rotate = 89.77] [fill={rgb, 255:red, 0; green, 0; blue, 0 }  ][line width=0.08]  [draw opacity=0] (5.36,-2.57) -- (0,0) -- (5.36,2.57) -- cycle    ;
\draw [line width=0.75]    (357.89,134.98) -- (357.68,122.55) ;
\draw [shift={(357.63,119.55)}, rotate = 89.06] [fill={rgb, 255:red, 0; green, 0; blue, 0 }  ][line width=0.08]  [draw opacity=0] (5.36,-2.57) -- (0,0) -- (5.36,2.57) -- cycle    ;
\draw [line width=0.75]    (366.95,134.98) -- (366.74,118) ;
\draw [shift={(366.7,115)}, rotate = 89.27] [fill={rgb, 255:red, 0; green, 0; blue, 0 }  ][line width=0.08]  [draw opacity=0] (5.36,-2.57) -- (0,0) -- (5.36,2.57) -- cycle    ;
\draw    (318.93,45.32) -- (374.5,45.32) ;
\draw [line width=0.75]    (346.71,45.32) -- (346.71,35.9) ;
\draw [shift={(346.71,32.9)}, rotate = 90] [fill={rgb, 255:red, 0; green, 0; blue, 0 }  ][line width=0.08]  [draw opacity=0] (5.36,-2.57) -- (0,0) -- (5.36,2.57) -- cycle    ;
\draw [line width=0.75]    (326.69,45.3) -- (326.69,26.53) ;
\draw [shift={(326.69,23.53)}, rotate = 90] [fill={rgb, 255:red, 0; green, 0; blue, 0 }  ][line width=0.08]  [draw opacity=0] (5.36,-2.57) -- (0,0) -- (5.36,2.57) -- cycle    ;
\draw [line width=0.75]    (332.1,45.32) -- (332.14,18) ;
\draw [shift={(332.14,15)}, rotate = 90.07] [fill={rgb, 255:red, 0; green, 0; blue, 0 }  ][line width=0.08]  [draw opacity=0] (5.36,-2.57) -- (0,0) -- (5.36,2.57) -- cycle    ;
\draw [line width=0.75]    (365.95,45.3) -- (365.74,19.55) ;
\draw [shift={(365.72,16.55)}, rotate = 89.54] [fill={rgb, 255:red, 0; green, 0; blue, 0 }  ][line width=0.08]  [draw opacity=0] (5.36,-2.57) -- (0,0) -- (5.36,2.57) -- cycle    ;
\draw [line width=0.75]    (370.37,45.32) -- (370.37,35.1) ;
\draw [shift={(370.37,32.1)}, rotate = 90] [fill={rgb, 255:red, 0; green, 0; blue, 0 }  ][line width=0.08]  [draw opacity=0] (5.36,-2.57) -- (0,0) -- (5.36,2.57) -- cycle    ;
\draw [color={rgb, 255:red, 0; green, 0; blue, 0 }  ,draw opacity=1 ]   (389.43,100.85) -- (497,100.85) ;
\draw [color={rgb, 255:red, 0; green, 0; blue, 0 }  ,draw opacity=1 ]   (170.5,55) -- (252,54.85) ;
\draw [color={rgb, 255:red, 0; green, 0; blue, 0 }  ,draw opacity=1 ]   (166.5,148) -- (248,147.85) ;
\draw  [color={rgb, 255:red, 74; green, 74; blue, 74 }  ,draw opacity=1 ][fill={rgb, 1:red, 0.7; green, 0.75; blue, 0.71 }  ,fill opacity=0.62 ] (75,109.6) .. controls (75,105.4) and (78.4,102) .. (82.6,102) -- (113.9,102) .. controls (118.1,102) and (121.5,105.4) .. (121.5,109.6) -- (121.5,132.4) .. controls (121.5,136.6) and (118.1,140) .. (113.9,140) -- (82.6,140) .. controls (78.4,140) and (75,136.6) .. (75,132.4) -- cycle ;
\draw  [color={rgb, 255:red, 74; green, 74; blue, 74 }  ,draw opacity=1 ][fill={rgb, 1:red, 0.7; green, 0.75; blue, 0.71 }  ,fill opacity=0.62 ] (79,22.6) .. controls (79,18.4) and (82.4,15) .. (86.6,15) -- (117.9,15) .. controls (122.1,15) and (125.5,18.4) .. (125.5,22.6) -- (125.5,45.4) .. controls (125.5,49.6) and (122.1,53) .. (117.9,53) -- (86.6,53) .. controls (82.4,53) and (79,49.6) .. (79,45.4) -- cycle ;

\draw (535,72) node [anchor=north west][inner sep=0.75pt]  [font=\normalsize]  {$D\{\cdot \}$};
\draw (612,69.4) node [anchor=north west][inner sep=0.75pt]  [font=\normalsize]  {$y(t)$};
\draw (99.56,121.79) node    {$\mathcal{C}_{K}(\cdot)$};
\draw (101.57,33.09) node    {$\mathcal{C}_{1}(\cdot)$};
\draw (89,62) node [anchor=north west][inner sep=0.75pt]  [font=\LARGE]  {$\vdots $};
\draw (26,62) node [anchor=north west][inner sep=0.75pt]  [font=\LARGE]  {$\vdots $};
\draw (15,25.4) node [anchor=north west][inner sep=0.75pt]  [font=\normalsize]  {$x_{1}( t)$};
\draw (13,113.4) node [anchor=north west][inner sep=0.75pt]  [font=\normalsize]  {$x_{K}( t)$};
\draw (162,6.4) node [anchor=north west][inner sep=0.75pt]  [font=\normalsize]  {$s_{1}( t)$};
\draw (161,98.4) node [anchor=north west][inner sep=0.75pt]  [font=\normalsize]  {$s_{K}( t)$};
\draw (192,62) node [anchor=north west][inner sep=0.75pt]  [font=\LARGE]  {$\vdots$};
\draw (284,21.4) node [anchor=north west][inner sep=0.75pt]  [font=\normalsize]  {$h_{1}( t)$};
\draw (272,119.9) node [anchor=north west][inner sep=0.75pt]  [font=\normalsize]  {$h_{K}( t)$};
\draw (430,19.4) node [anchor=north west][inner sep=0.75pt]  [font=\normalsize]  {$v(t)$};


\begin{axis}[
width=4cm,
height=3.5cm,
xticklabels=none,
 yticklabels=none,
 xtick=\empty,
 ytick=\empty,
 axis line style={draw=none},
 xshift=4.4cm,yshift=-0.5cm,
 ]
\addplot[
    domain=-20:20,
    samples=500, 
    color=copperrose,
]{0.1*(-5*sin(1.3*deg(x))/x)};
\end{axis}

\begin{axis}[
width=4cm,
height=3.5cm,
xticklabels=none,
 yticklabels=none,
 xtick=\empty,
 ytick=\empty,
 axis line style={draw=none},
 xshift=4.4cm,yshift=2cm,
 ]
\addplot[
    domain=-20:20,
    samples=500, 
    color=ceil,
]{0.1*(-5*sin(1.1*deg(x-5))/(x-5))};
\end{axis}
\begin{axis}[
width=4.8cm,
height=3.5cm,
xticklabels=none,
 yticklabels=none,
 xtick=\empty,
 ytick=\empty,
 axis line style={draw=none},
 xshift=10cm,yshift=0.75cm,
 ]
\addplot[
    domain=-20:20,
    samples=500, 
    color=cerulean,
]{-15*sin(deg(x+14))/(x+14)+cos(deg(x)-1)-10*sin(deg(x-7))/(x-7)+cos(deg(x)-1) -20*sin(deg(x-13))/(x-13) };
\end{axis}

\end{tikzpicture}

}
    
    \caption{An illustration of the mathematical model of \ac{GB2D} problem. Every user transmits waveform $s_k(t)$ over channel $h_k(t)$ involves $P_k$ multi-path. Afterward, the sum of all these convolved signals is received by $v(t)$.   }
    \label{fig:system}
\end{figure*}
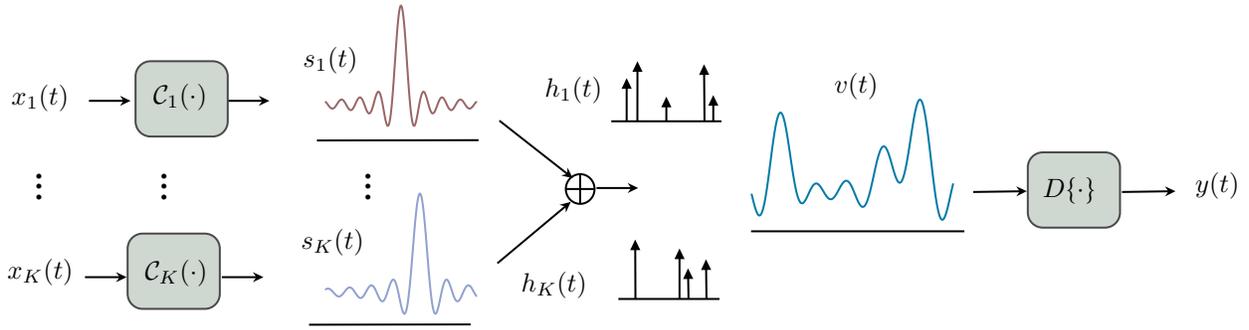


To facilitate explanation, we consider a scenario where multiple users transmit messages through multiple frequency-selective channels towards a central \ac{BS} (as described in \cite[Eq. 19]{sayeed2010wireless}). The \ac{BS} receives a combined signal comprising contributions from all users, which is then processed through a sensing filter (see Fig. \ref{fig:system}). The goal is to simultaneously estimate the transmitted messages and channels from the received measurements at the \ac{BS}, which is a challenging nonlinear problem. 

This scenario appears in a variety of applications, including over-the-air computation~\cite{SaeedBlind2022,razavikia2023computing}, super-resolution single-molecule imaging~\cite{three-dimensionalsuper,SaeedBinary2020,razavikia2019sampling,maskan2023demixing,valiulahi2019two},  multi-user multipath channel estimation~\cite{jung2017blind,daei2023blind}, blind calibration in multi-channel sampling systems~\cite{vetterli2010multichannel,SayyariBlind2021}, random access~\cite{daei2023blinda} and integrated (radar) sensing and communications~\cite{liu2020joint,seidi2022novel,safari2021off}.

\subsection{Related work}

The problem of recovering messages and channels in the model described above falls into the class of blind deconvolution techniques used to solve inverse problems. These techniques have made notable progress in addressing blind deconvolution problems, with a focus on sparse signals consisting of a single user \cite{jain2017non,ahmed2013blind,ling2015self,candes2013phaselift,daei2019distribution}. The conventional method involves assuming that the continuous channel parameters lie on a predefined domain of grids, which can be estimated using well-established methods such as $\ell_1$ minimization. However, the predefined grids may not accurately match the true continuous-index values of parameters, leading to basis mismatch issues that can degrade the performance of blind deconvolution. To address these issues, recent work has focused on continuous parameter estimation, with a blind deconvolution method developed in \cite{chi2016guaranteed} to recover continuous parameters in the case of a single user.

In this work, we tackle a challenging and more generalized model that involves a mixture of blind deconvolution problems, with transmitted signals being encoded with different codebooks and arbitrary channels that do not follow a specific channel model. Further, we consider the availability of a general sensing filter at the \ac{BS} that allows us to observe compressed linear combinations of data samples instead of the whole samples, as in \cite{Jacome2020DualBlind}, to ensure practicality and generality. It is noteworthy that our proposed method is deterministic and is independent of either the channel or message distribution, which sets it apart from existing statistical techniques (such as approximate message passing~\cite{ke2020compressive}) that can only work on specific distributions. Finally, we prove a remarkable result that when all users use the same codebook to transmit their messages, our proposed optimization problem is independent of the total number of users.   

\subsection{Contributions}

We propose a novel optimization framework, named \ac{GB2D}, that leverages specific features of the channels and transmitted signals. Specifically, each user's channel has few dominant scattering paths, and each user employs a distinct channel coding scheme. \ac{GB2D} utilizes a lifting technique to convert the demixing of nonlinear problems into high-dimensional matrices containing continuous channel parameters, which transforms multiple features of the channels and transmitted signals into a single specific feature of a matrix tuple lying in a higher dimension. A tractable convex optimization problem is proposed to recover the continuous channel parameters by promoting the specific feature of the matrix tuple, followed by a least-squares problem to estimate the transmitted messages. In addition, this work provides conditions under which the solution to \ac{GB2D} is unique and optimal, and simulation results demonstrate its effectiveness in recovering the channel parameters and transmitted messages with a sufficient number of samples.

Specifically, our contributions are summarized as follows:

\begin{itemize}
    \item \textbf{Blind message recovery and channel estimation with any linear encoder and sensing filter}:  \ac{GB2D} recovers the messages and channels without spending training resources and provides a general communication framework where each user employs a distinct codebook, and the \ac{BS} employs a linear filter modeling matched filter or sensing block of a communication system. 
    \item \textbf{Independent of the number of users}: In a case wherein all users use the same codebook to transmit their messages, the proposed optimization problem is independent of the total number of users.
    \item \textbf{Tractable complexity}: We propose a tractable convex optimization problem to recover continuous channel parameters by promoting the specific feature of the channels and transmitted signals and then recovering the messages. 
    \item  \textbf{Optimality condition}: We provide a theoretical guarantee that the solution to \ac{GB2D} is unique and optimal under some minimum separation conditions on the multipath channel delays. 
    \item \textbf{Message and channel distribution are arbitrary}: In contrast to the statistical channel estimation methods, e.g.,covariance-based methods and approximate message passing,  \ac{GB2D} does not require any assumptions for the distributions of users' messages and channels.
\end{itemize}

The rest of the paper is organized as follows.  In Section~\ref{sec.model}, the problem formulation is formalized. In Section~\ref{sec.proposed}, we provide the \ac{GB2D} method, which includes the convex optimization and dual problem to localize the spikes. Section~\ref{sec.simulations} verifies the performance of \ac{GB2D} through simulations. Lastly, the paper is concluded in Section~\ref{sec.conclusion}.

The paper shows vectors and matrices in boldface lower-and upper-case letters, respectively. We use $\bm{X}^{\dagger}$ to show the pseudo inverse of matrix $\bm{X}$.  For vector $\bm{x}\in\mathbb{C}^N$ and matrix $\bm{X}\in\mathbb{C}^{N_1\times N_2}$, the norms $\ell_2$ is defined as $\|\bm{x}\|_2:= \sqrt{\sum\nolimits_{i=1}^n|x(i)|^2}$. 
We further use $\odot$ to show Hadamard products (element-wise products). The Toeplitz lifting operator $\mathscr{T}: \mathbb{C}^N \to \mathbb{C}^{N\times N}$ for given vector $\bm{x}$  is defined as follows
\begin{equation}
    \label{eq:Hankel}
    \mathscr{T}(\bm{x}): = \begin{bmatrix}
        x_{1} & x_{2} & \dots & x_{N} \\
        \Bar{x}_{2} & x_{1} & \dots & x_{N-1} \\
        \vdots \\
        \Bar{x}_{N} & \Bar{x}_{N-1} & \dots & x_{1} 
    \end{bmatrix}, 
\end{equation}
where the $(i,j)$-th  element is given by $[\mathscr{T}(\bm{x})]_{i,j} = x_{i-j+1}$ for $i\geq j$ and $[\mathscr{T}(\bm{x})]_{i,j} = \Bar{x}_{i-j+1}$ for $i<j$. Also $\bm{e}_n$ stands for the $n$-th column of identity matrix $\bm{I}_{N}$. We use $\langle \cdot, \cdot \rangle$ to show the inner product operator where for two arbitrary matrices $\bm{A}, \bm{B}$, i.e., $\langle \bm{A}, \bm{B}\rangle$ represents ${\rm Tr} (\bm{B}^{\mathsf  H}\bm{A})$ and for two continuous functions $f(t)$ and $g(t)$ it means $\int_{-\infty}^{\infty}f(t)g(t)dt$ by $\langle f(t), g(t) \rangle$ based on context. The notation $\bm{A}\succeq \bm{0}$ means $\bm{A}$ is a positive semidefinite matrix.

 \section{System Model and Problem Formulation}\label{sec.model}

For the purpose of exposition, we consider $K$ single-antenna users transmitting their messages toward a \ac{BS} over the \ac{MAC}.  Let  $x_k(t)$ be the message of user $k$; it is mapped into modulated band-limited signal $s_k(t)$ using linear encoder $\mathcal{C}_k(\cdot)$, i.e., $s_k(t):= \mathcal{C}_k(x_k(t))$ for transmitting over the \ac{MAC}. Then, all users transmit simultaneously over the same frequency or codes, and the \ac{BS} records the signal $v(t)$, which is the sum of all these convolved signals given by 
\begin{align}\label{eq:mymodel}
v(t)=\sum_{k=1}^{K}h_k(t)\circledast s_k(t),
\end{align}
where $h_k(t)=\sum_{\ell=1}^{P_k}g_{\ell}^k\delta(t-\overline{\tau}_{\ell}^k)$, is the impulse response of the frequency selective channel corresponding to user $k$, and $\circledast$ is the convolution operator. Furthermore, $P_k$, $\overline{\tau}_{\ell}^k$, and $g_{\ell}^k$ are the number of multipath delays, the delay, and the complex amplitude of the communication path $k$ corresponding to user $k$, respectively. The channel delays $\overline{\tau}_{\ell}^k$s can take any arbitrary continuous values in $[0, T)$ in which $T$ denotes the duration of the observation time.  Afterward, the convoluted signal $v(t)$ goes through linear system $\mathcal{D}(\cdot)$ (e.g., matched-filter or low pass filter) whose output becomes signal $y(t)$, i.e.,  
     \begin{align}\label{eq:measure}
     y(t) = \mathcal{D}\{v(t)\},  \quad t \in [0,T].
     \end{align}      
Then, after sampling, the measurements are given by $y_m:= y(\frac{m}{T})$ for $m\in [M]$. The goal is to estimate the set of channel delays and amplitudes of $\{h_k(t)\}_{k=1}^K$, as well as the unknown transmitted messages $\{x_k(t)\}_{k=1}^K$ from the available measurements  $y_1, \ldots, y_M$ at the \ac{BS} (see Fig~\ref{fig:system}). We refer to the solution to this problem as \ac{GB2D}.

\subsection{Problem formulation}

Assuming the measured signal $y(t)$ is square-integrable in Lebesgue’s sense (or Band-limited), we can expand it as $y(t):=  \sum\nolimits_{m=1}^M y_m \varphi_m(t)$  where  $\varphi_n$s  are compact support basis functions for $t\in [0, T]$ that are orthonormal, i.e., 
 \begin{align}
    \langle\varphi_i(t), \varphi_j(t) \rangle :=  \int_{-\infty}^{\infty}\varphi_i(t)\varphi_j(t) dt  = \delta_{i-j},     
 \end{align}
 where $\delta_{i-j}$ is the discrete delta Dirac function. Then, $m$-th sample of the signal in~\eqref{eq:measure}, can be written as 
 \begin{align}
    \label{eq:innerVL}
     \nonumber y_m &= \langle \mathcal{D}\{v(t)\}, \varphi_m \rangle  \\ \nonumber
     & = \langle v(t), \underbrace{\mathcal{D}^*\{\varphi_m \}}_{d_{m}(t)}\rangle \\ 
     \nonumber & = \langle \mathscr{F}^*\mathscr{F}\{v(t)\}, d_{m}(t)\rangle \\ & = \nonumber \langle \mathscr{F}\{v(t)\}, \mathscr{F}\{\ell_{m}(t)\}\rangle \\
     & = \langle V(f), D_m(f)\rangle,
 \end{align}
where $\mathscr{F}(\cdot)$ denotes linear Fourier operator, and $V(f)$ and $D_m(f)$ are the Fourier transform of $v(t)$ and $d_m(t)$, respectively. Let $\{s_k(t)\}_{k=1}^K$ be the transmitted waveforms whose spectrum lie in the interval $[-B,B]$. Taking the Fourier transform of~\eqref{eq:mymodel}; we have the following.
 \begin{align} \label{eq.Yf}
V(f)=\sum_{k=1}^{K}H_k(f)S_k(f), ~\forall f\in [-B,B],
 \end{align}   
 where $V(f)$, $H_k(f)$, and $S_k(f)$ are the Fourier transform of $v(t)$, $h_k(t)$, and $s_k(t)$, respectively. Substituting \eqref{eq.Yf} into \eqref{eq:innerVL}, we obtain
  \begin{align}
     \nonumber y_m = \sum_{k=1}^{K} \langle H_k(f)S_k(f), D_m(f)\rangle.
 \end{align}
 By uniformly sampling \eqref{eq.Yf} at $N$ points $f_n=Bn/\lfloor (N-1)/2\rfloor,~n= -\lceil (N-1)/2\rceil, \ldots, \lfloor (N-1)/2\rfloor$ or $n=0,\ldots, N-1$ provided that $BT\le \lfloor (N-1)/2\rfloor$, we reach
 \begin{align}\label{eq.sampledmodel1}
  y_m =  \langle \bm{d}_m, \sum_{k=1}^K \bm{h}_k\odot \bm{s}_k \rangle = \langle \bm{d}_m, \bm{v} \rangle,
 \end{align}
where vectors $\bm{h}_k, \bm{s}_k, \bm{d}_m$ and  $\bm{v}$ are defined as 
\begin{align*}
    \bm{h}_k & = [H(f_1),\ldots,H(f_N)]^{\mathsf{T}}, \\
     \bm{s}_k & = [S(f_1),\ldots,S(f_N)]^{\mathsf{T}}, \\ 
     \bm{d}_m & = [D_m(f_1),\ldots,D_m(f_N)]^{\mathsf{T}}, \\
     \bm{v} & =  [V(f_1),\ldots,V(f_N)]^{\mathsf{T}}.
\end{align*}
Note that to set $N$ as small as possible without loss of generality, we choose $N= 2BT + 1$.
The relation in \eqref{eq.sampledmodel1} can be represented in a matrix form as
 \begin{align}
 \bm{y}=\bm{D}\sum_{k=1}^K \bm{h}_k\odot \bm{s}_k,
 \end{align}
 where $\bm{y}=[y_{1}, \ldots,y_{M}]^{\mathsf{T}}$ and $\bm{D} = [\bm{d}_1,\ldots, \bm{d}_M]^{\mathsf{T}} \in \mathbb{C}^{M\times N}$. Now, recall that $h_k(t) = \sum_{\ell=1}^{P_k}g_{\ell}^k\delta(t-\tau_{\ell}^k)$ and $s_k(t)= \mathcal{C}_k(x_k(t))$ where ${\tau}_{\ell}^k:=  \overline{\tau}_{\ell}^k/T$, then we can write  
\begin{align}
    \label{eq.sampledmodel2}
   \bm{y} = \bm{D}\sum_{k=1}^K \sum_{\ell=1}^{P_k}g_{\ell}^k \bm{a}(\tau_{\ell}^k)\odot \bm{C}_k\bm{x}_k,
\end{align}
 where $\bm{a}(\tau):=[1,{\rm e}^{-j2\pi\tau}, \ldots ,{\rm e}^{-j2\pi(N-1)\tau}]^{\mathsf{T}}$ and 
\begin{align}\label{eq.subspace_assumption}
 \bm{s}_k=\bm{C}_k\bm{x}_k.
 \end{align} 
  Also, $\bm{C}_k:=[\bm{c}_{1}^k, \ldots, \bm{c}_{N}^k]^{\mathsf{T}}\in\mathbb{C}^{N\times M_k}$ is codebook matrix corresponding to encoder $\mathcal{C}_k$ which is a known basis of the subspace with $N\gg M_k$. Further, $\bm{x}_k\in\mathbb{C}^{M_k}$ is the Fourier transform of message vector of user $k$.  Without loss of generality, we assume that the energy of the message signal is normalized, i.e., $\|\bm{x}_k\|_2=1$ for $k \in [K]$. Our goal is to recover $\tau_{\ell}^k$s, $g_{\ell}^k$s, and $\bm{x}_k$s from the observation vector $\bm{y}\in\mathbb{C}^{M}$. Note that it is unavoidable to
have phase ambiguities for recovering $\bm{x}_k$'s and $\bm{h}_k$'s because of any $\alpha_k\in\mathbb{C}\setminus\{0\}$, we have
 \begin{align}
 \bm{y}= \bm{D}\sum_{k=1}^K\alpha_k\bm{h}_k\odot \bm{C}_k\frac{\bm{x}_k}{\alpha_k}.
 \end{align}

In the next section, we present the \ac{GB2D} method for demixing the measured signals by solving a convex optimization.

 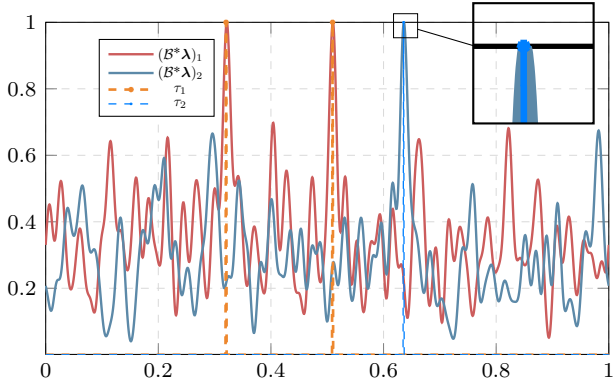
\begin{figure}[!t]
\centering
 \begin{tikzpicture} 

 \begin{scope}[spy using outlines={rectangle, magnification=5,
   width=1.6cm,height=1.6cm,connect spies}]
    \begin{axis}[
        width=0.5\textwidth,
        height=6cm,
        xmin=0, xmax=1,
        ymin=1e-3, ymax=1,
        legend style={nodes={scale=0.55, transform shape}, at={(0.3,0.95)}}, 
        ticklabel style = {font=\footnotesize},
        ymajorgrids=true,
        xmajorgrids=true,
        grid style=dashed,
        grid=both,
        grid style={line width=.1pt, draw=gray!10},
        major grid style={line width=.2pt,draw=gray!30},
    ]
    \addplot[ smooth,
             thin,
        color=chestnut,
        line width=0.9pt,
        ]
    table[x=t,y=D1]
    {Data/PolyDual3.dat};
\addplot[ smooth,
             thin,
        color=airforceblue,
        line width=0.9pt,
        ]
    table[x=t,y=D2]
    {Data/PolyDual3.dat};
    \addplot[ 
        color=cadmiumorange,
        mark=star,
        dashed,
        line width=1pt,
        mark size=1pt,
        ]
    table[x=t,y=X1]
    {Data/PolyDual3.dat};
    \addplot[ 
        color=azure,
        mark=star,
        dashed,
        line width=0.5pt,
        mark size=0.5pt,
        ]
    table[x=t,y=X2]
    {Data/PolyDual3.dat};
    \legend{$(\mathcal{B}^*\bm{\lambda})_1$, $(\mathcal{B}^*\bm{\lambda})_2$,$\tau_1$,$\tau_2$};
     \path (0.639, 0.99) coordinate (X);
    \end{axis}
    \spy [black] on (X) in node (zoom) [left] at ([xshift=2.5cm,yshift=-0.5cm]X);
    \end{scope}
\end{tikzpicture}

  \caption{Delay estimation via dual polynomials with order  $N=64$ for $K=2$ and $P_1=2,P_2=1$ with $M_1=M_2 =5$.  }
   \label{fig:Daul}
\end{figure}


\section{Proposed Method}\label{sec.proposed}
 In this section, we introduce the main idea \ac{GB2D} method and propose a convex optimization to recover all the channel parameters ${\tau}_{\ell}^k$'s, $\ell \in [P_k]$ and transmitted waveforms $s_k(t)$s with general known codebook matrices $\bm{C}_k$'s.  Invoking the subspace assumption \eqref{eq.subspace_assumption},  $n$-th Fourier samples of signal $v(t)$ can be written as 
 \begin{align}\label{eq.sampledmodel3}
 V(f_n)&= \sum_{k=1}^K\sum_{\ell=1}^{P_k}g_{\ell}^k\bm{e}_n^{\mathsf{T}}\bm{a}(\tau_{\ell}^k)\bm{x}_k^{\mathsf{T}}\bm{c}^k_n,
 \end{align}
 where $\bm{e}_n$ stands for the $n$-th column of $\bm{I}_{N}$. Let $\bm{X}_k=\sum_{\ell=1}^{P_k} g_{\ell}^k\bm{x}_k\bm{a}(\tau_{\ell}^k)^{\mathsf{T}}\in\mathbb{C}^{M_k\times N}$. Using the lifting trick \cite{ling2015self}, the measurements $V(f_n), n\in [N]$ in \eqref{eq.sampledmodel3} can be written as
 \begin{align*}
 &V(f_n)=\sum_{k=1}^K\sum_{\ell=1}^{P_k}g_{\ell}^k {\rm Tr}\Big(\bm{c}^k_n\bm{e}_n^{\mathsf{T}}\bm{a}(\tau_{\ell}^k)\bm{x}_k^{\mathsf{T}}\Big)=\sum_{k=1}^K\big\langle \bm{X}_k, \bm{c}^k_n\bm{e}_n^{\mathsf{T}}\big\rangle.
 \end{align*} 
 Writing in matrix form, we have $\bm{v}=\mathcal{C}(\bm{\mathcal{X}}),$ where $\bm{\mathcal{X}}:=(\bm{X}_k)_{k=1}^K\in \bigoplus_{k=1}^k\mathbb{C}^{M_k\times N}$ is the matrix tuple of interest and $\mathcal{C}$ is the linear measurement mapping defined as
 \begin{align*}
 \bigoplus_{k=1}^K\mathbb{C}^{M_k\times N}\rightarrow \mathbb{C}^N,~~ \bm{\mathcal{Z}}\rightarrow \Bigg(\sum_{k=1}^K\big\langle \bm{Z}_k, \bm{c}^k_n\bm{e}_n^{\mathsf{T}}\big\rangle\Bigg)_{n=1}^{N}.	
 \end{align*}
Then, by defining $ \mathcal{B} := \bm{D}\mathcal{C}$, the measurements $\bm{y}$ reads to
\begin{align}\label{eq:measure_model1}
    \bm{y}  =  \mathcal{B}(\bm{\mathcal{X}}). 
\end{align} 
 
 In model~\eqref{eq.sampledmodel3}, the number of delays $\{P_k\}_{k=1}^K$ (e.g., in a multipath channel in multi-user wireless systems) is small. 
 Thus, we define the atomic norm~\cite{chandrasekaran2012convex}
 \begin{align}\label{eq.atomic_def}
 &\|\bm{Z}\|_{\mathcal{A}_k}:=\inf\{t>0: \bm{Z}\in t{\rm conv}(\mathcal{A}_k)\}\nonumber\\
 &=\inf_{\substack{c_{\ell}, \tau_{\ell}\\\|{\bm{x}_k}\|_2=1}}\Big\{\sum_{\ell}|c_{\ell}|:~\bm{Z}=\sum_{\ell}c_{\ell} \bm{x}_k\bm{a}(\tau_{\ell})^{\mathsf{T}}  \in\mathbb{C}^{M_k\times N} \Big\}
 \end{align}
 associated with the atoms
 \begin{align*}
 \mathcal{A}_k=\big\{\bm{x}\bm{a}(\tau)^{\mathsf{T}}: \tau\in[0,1), \|\bm{x}\|_2=1, \bm{x}\in\mathbb{C}^{M_k} \big\}, k\in [K].
 \end{align*}
  The atomic norm $\|\bm{X}_k\|_{\mathcal{A}_k}$ can be regarded as the best convex alternative for the smallest number of atoms $\mathcal{A}_k$ needed to represent a signal $\bm{X}_k$. Hence, we are interested in recovering the matrix tuple $\bm{\mathcal{X}}:=(\bm{X}_k)_{k=1}^K$ by motivating its atomic sparsity by solving the following optimization problem. 
 \begin{align}\label{eq.primalprob}
 \min_{\bm{\mathcal{Z}}=(\bm{Z}_k)_{k=1}^K} ~\sum_{k=1}^{K}\|\bm{Z}_k\|_{\mathcal{A}_k}\quad 
 \bm{y}_{M\times 1}=\mathcal{B}(\bm{\mathcal{Z}}).	
 \end{align} 
Finding the optimal parameters in \eqref{eq.primalprob} is not an easy task because it involves an infinite-dimensional variable optimization due to the continuity of the set. Alternatively, we can solve the dual problem explained in the next section. 

In what follows, we state the conditions for the uniqueness of the solution to optimization in \eqref{eq.primalprob} and the exact recovery of channel parameters and message data. Before expressing Proposition \ref{prop.optimality}, we define the separation between the delays of channel $k$ as
 \begin{align}
 \Delta_k:=\min_{\ell \neq q}|\tau^k_{\ell}-\tau^k_q|,
 \end{align} 
and the minimum separation between all users by $\Delta:=\min_{i} \Delta_k$. The absolute value in the latter definition is evaluated as the wrap-around distance on the unit circle.

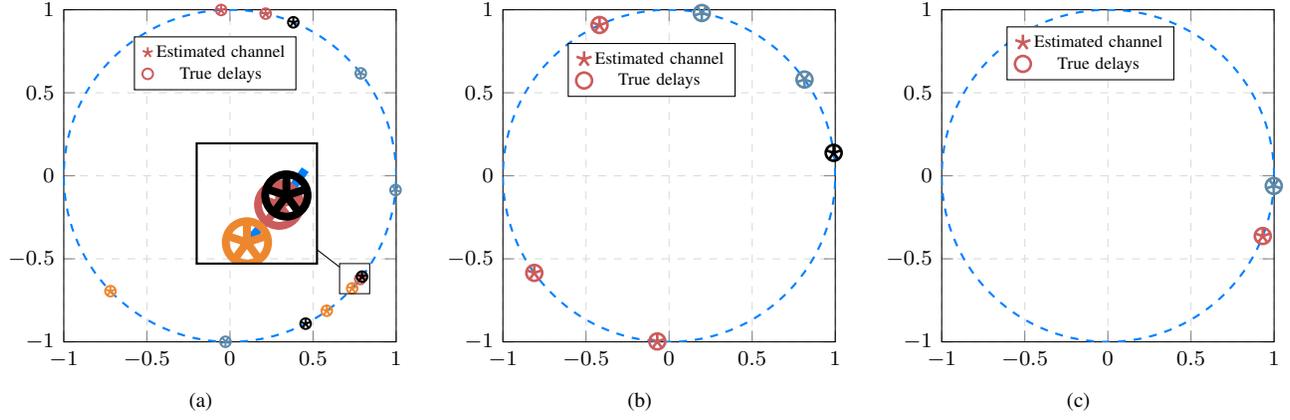
\begin{figure*}[!t]
\centering
\subfigure[]{\label{Fig(a):Circ}
 \begin{tikzpicture} 
 \begin{scope}[spy using outlines={rectangle, magnification=4,
   width=1.6cm,height=1.6cm,connect spies}]
    \begin{axis}[
        width=6cm,
        height=6cm,
        xmin=-1, xmax=1,
        ymin=-1, ymax=1,
        legend style={nodes={scale=0.65, transform shape}, at={(0.7,0.92)}}, 
        ticklabel style = {font=\footnotesize},
        ymajorgrids=true,
        xmajorgrids=true,
        grid style=dashed,
        grid=both,
        grid style={line width=.1pt, draw=gray!10},
        major grid style={line width=.2pt,draw=gray!30},
    ]
    \addplot[only marks,
        color=chestnut,
        mark=star,
        line width=0.75pt,
        mark size=2pt,
        ]
    table[x=X1,y=Y1]
    {Data/Sim1.dat};
   \addplot[only marks,
        color=chestnut,
        mark=o,
        line width=0.75pt,
        mark size=2pt,
        ]
    table[x=TX1,y=TY1]
    {Data/Sim2.dat};

    \draw[azure,thick,dashed] (axis cs:0,0) circle [radius=1];
    
    \addplot[only marks,
        color=airforceblue,
        mark=star,
        mark options = {rotate = 180},
        line width=0.75pt,
        mark size=2pt,
        ]
    table[x=X2,y=Y2]
    {Data/Sim1.dat};
    \addplot[only marks,
        color=airforceblue,
        mark=o,
        line width=0.75pt,
        mark size=2pt,
        ]
    table[x=TX2,y=TY2]
    {Data/Sim2.dat};
    \addplot[only marks,
        color=black,
        mark=star,
        line width=0.75pt,
        mark size=2pt,
        ]
    table[x=X3,y=Y3]
    {Data/Sim1.dat};
    \addplot[only marks,
        color=black,
        mark=o,
        line width=0.75pt,
        mark size=2pt,
        ]
    table[x=TX3,y=TY3]
    {Data/Sim2.dat};
    \addplot[only marks,
        color=cadmiumorange,
        mark=star,
        line width=0.75pt,
        mark size=2pt,
        ]
    ttable[x=X4,y=Y4]
    {Data/Sim1.dat};
    \addplot[only marks,
        color=cadmiumorange,
        mark=o,
        line width=0.75pt,
        mark size=2pt,
        ]
    table[x=TX4,y=TY4]
    {Data/Sim2.dat};
    \legend{Estimated channel, True delays};
     \path (0.75, -0.62) coordinate (X);
    \end{axis}
    \spy [black] on (X) in node (zoom) [left] at ([xshift=-0.5cm,yshift=1cm]X);
    \end{scope}
\end{tikzpicture}
}
\subfigure[]{\label{Fig(b):Circ}
\begin{tikzpicture} 
    \begin{axis}[
        width=6cm,
        height=6cm,
        xmin=-1, xmax=1,
        ymin=-1, ymax=1,
        legend style={nodes={scale=0.65, transform shape}, at={(0.7,0.9)}}, 
        ticklabel style = {font=\footnotesize},
        ymajorgrids=true,
        xmajorgrids=true,
        grid style=dashed,
        grid=both,
        grid style={line width=.1pt, draw=gray!10},
        major grid style={line width=.2pt,draw=gray!30},
    ]
    \addplot[only marks,
        color=chestnut,
        mark=star,
        line width=1pt,
        mark size=3pt,
        ]
    table[x=TEX1,y=TEY1]
    {Data/resultR3L5N128NEW.dat};
   \addplot[only marks,
        color=chestnut,
        mark=o,
        line width=1pt,
        mark size=3pt,
        ]
    table[x=TX1,y=TY1]
    {Data/resultR3L5N128NEW.dat};

    \draw[azure,thick,dashed] (axis cs:0,0) circle [radius=1];
    
    \addplot[only marks,
        color=airforceblue,
        mark=star,
        mark options = {rotate = 180},
        line width=1pt,
        mark size=3pt,
        ]
    table[x=TEX2,y=TEY2]
    {Data/resultR3L5N128NEW.dat};
    \addplot[only marks,
        color=airforceblue,
        mark=o,
        line width=1pt,
        mark size=3pt,
        ]
    table[x=TX2,y=TY2]
    {Data/resultR3L5N128NEW.dat};
    \addplot[only marks,
        color=black,
        mark=star,
        line width=1pt,
        mark size=3pt,
        ]
    table[x=TEX3,y=TEY3]
    {Data/resultR3L5N128NEW.dat};
    \addplot[only marks,
        color=black,
        mark=o,
        line width=1pt,
        mark size=3pt,
        ]
    table[x=TX3,y=TY3]
    {Data/resultR3L5N128NEW.dat};
    \legend{Estimated channel, True delays};
    \end{axis}
\end{tikzpicture}
}
\subfigure[]{\label{Fig(c):Circ}
\begin{tikzpicture} 
    \begin{axis}[
        width=6cm,
        height=6cm,
        xmin=-1, xmax=1,
        ymin=-1, ymax=1,
        legend style={nodes={scale=0.65, transform shape}, at={(0.7,0.95)}}, 
        ticklabel style = {font=\footnotesize},
        ymajorgrids=true,
        xmajorgrids=true,
        grid style=dashed,
        grid=both,
        grid style={line width=.1pt, draw=gray!10},
        major grid style={line width=.2pt,draw=gray!30},
    ]
    \addplot[only marks,
        color=chestnut,
        mark=star,
        line width=1pt,
        mark size=3pt,
        ]
    table[x=X1,y=Y1]
    {Data/resultR2L16N64NEW.dat};
   \addplot[only marks,
        color=chestnut,
        mark=o,
        line width=1pt,
        mark size=3pt,
        ]
    table[x=TX1,y=TY1]
    {Data/resultR2L16N64NEW.dat};

    \draw[azure,thick,dashed] (axis cs:0,0) circle [radius=1];
    
    \addplot[only marks,
        color=airforceblue,
        mark=star,
        mark options = {rotate = 180},
        line width=1pt,
        mark size=3pt,
        ]
    table[x=X2,y=Y2]
    {Data/resultR2L16N64NEW.dat};
    \addplot[only marks,
        color=airforceblue,
        mark=o,
        line width=1pt,
        mark size=3pt,
        ]
    table[x=TX2,y=TY2]
    {Data/resultR2L16N64NEW.dat};
    \legend{Estimated channel, True delays};
    \end{axis}
\end{tikzpicture}
}
  \caption{Performance of \ac{GB2D}. Fig~\ref{Fig(a):Circ} shows the channel estimation for the case where for $K=4$ users and $P_1=\cdots=P_4=3$, form $N=200$ samples and the message size $M_k=5$ for $k\in [4]$. Fig~\ref{Fig(b):Circ} depict the performance of \ac{GB2D} for $K=3$ with $P_1=3,P_2=2,P_3 = 1$ from $N=128$ samples. Fig~\ref{Fig(c):Circ} shows  the output of \ac{GB2D} for the case $M_k=16$ and $M=64$. Each signal is shown with a different color code. 
  }
   \label{fig:Plor}
\end{figure*}


\begin{prop}\label{prop.optimality}
Denote the set of multipath's delays of $h_k(t)$ as $\mathcal{P}_k:=\{\tau_{\ell}^k\}_{\ell=1}^{P_k}$. The solution $\widehat{\bm{\mathcal{X}}}=(\widehat{\bm{X}}_k)_{k=1}^K$ of \eqref{eq.primalprob} is unique if  $\Delta \geq \frac{1}{N}$ and there exists a vector $\bm{\lambda}=[\lambda_{1}, \ldots, \lambda_{N}]^{\mathsf{T}}\in\mathbb{C}^N$ such that the vector-valued dual polynomials 
\begin{align}\label{eq.Qi}
\bm{q}_k(\tau)=(\mathcal{B}^*\bm{\lambda})_k\bm{a}^*(\tau)= \sum_{n=1}^{N}\lambda_n {\rm e}^{j2\pi n \tau} \bm{c}_n^k\in\mathbb{C}^{r_k}, 
\end{align}
for $k \in [K]$, satisfy the conditions
\begin{align}
&\bm{q}_k(\tau_{\ell})={\rm sgn}({c_{\ell}^{k}})\bm{x}_k~\forall \tau_{\ell}\in \mathcal{P}_k,~ k\in [K]\label{eq.supp_cond}\\
&\|\bm{q}_k(\tau)\|_2<1~~\forall \tau \in [0,1)\setminus \mathcal{P}_k,~ k\in [K].\label{eq.offsupp_cond}
\end{align}
 \end{prop}
\begin{proof}
See Appendix \ref{proof.optimality}.
\end{proof}

\begin{corl}
The dual polynomial $\bm{q}_k(\tau)$ only depends on its corresponding codebook, i.e.,   $\bm{C}_k$. Therefore, in the case where all users employ the same codebook matrix, all users have a common subspace, i.e., $\bm{C}_k = \bm{C} \in \mathbb{C}^{N\times M'}$ where $M'$ denotes message size for all the users,  all the atomic norms in \eqref{eq.primalprob} can be replaced with only one atom and its dual polynomial. 
\end{corl}            

 \subsection{Channel estimation via Dual Problem}
Before proceeding with the dual problem, let us define the dual polynomial of the atomic norm in the sequel.  In particular, the dual atomic norm $\|\cdot\|_{\mathcal{A}_k}^{\mathsf{d}}$ at an arbitrary point $\bm{Z}\in \mathbb{C}^{M_k\times N}$ is defined as
\begin{align}\label{eq.dualnorm1}
\|\bm{Z}\|_{\mathcal{A}_k}^{\mathsf{d}}&:=\sup_{\|\bm{X}\|_{\mathcal{A}_k}\le 1}{\rm Re}\big\{\langle \bm{Z}, \bm{X} \rangle\big\}\nonumber\\
&=\sup_{\substack{\tau\in [0,1)\\\|\bm{x}\|_2=1}}{\rm Re}\big\{\langle \bm{Z}, \bm{x}\bm{a}(\tau)^{\mathsf{T}} \rangle\big\} \nonumber\\
&=\sup_{\substack{\tau\in [0,1)\\ \|\bm{x}\|_2=1}}{\rm Re}\big\{\langle \bm{x}, \bm{Z}\bm{a}^*(\tau) \rangle\big\}\nonumber \\ 
& =\sup_{\tau\in [0,1)}\|\bm{Z}\bm{a}^*(\tau)\|_2.
\end{align}
Then, by assigning the Lagrangian vector $\bm{\lambda}\in\mathbb{C}^N$ to the equality constraint of \eqref{eq.primalprob}, we have
\begin{align}\label{eq.lagran1}
&L(\bm{\mathcal{Z}},\bm{\lambda})=\inf _{\bm{\mathcal{Z}}\in\bigoplus_{k=1}^K\mathbb{C}^{M_k\times N}}\Big[\sum_{k=1}^K\|\bm{Z}_k\|_{\mathcal{A}_k}+\langle \bm{\lambda},\bm{y}-\mathcal{B}(\bm{\mathcal{Z}}) \rangle\Big]\nonumber\\
&\langle \bm{\lambda},\bm{y}\rangle+\sum_{k=1}^K\inf_{\bm{Z}_k\in\mathbb{C}^{M_k\times N}}\Big[\|\bm{Z}_k\|_{\mathcal{A}_k}-\langle (\mathcal{B}^*\bm{\lambda})_k, \bm{Z}_k \rangle\Big],
\end{align}
where we used that $\langle \mathcal{B}^*\bm{\lambda}, \bm{\mathcal{Z}}\rangle=\sum_{k=1}^K\langle (\mathcal{B}^*\bm{\lambda})_k, \bm{Z}_k\rangle.$
By using H\"{o}lder's inequality, \eqref{eq.lagran1} becomes equivalent to
\begin{align}
\nonumber L(\bm{\mathcal{Z}},\bm{\lambda})&=\langle  \bm{\lambda},\bm{y}\rangle
\\
& \nonumber+\sum_{k=1}^K\inf_{\bm{Z}_k\in\mathbb{C}^{M_k\times N}}\Big[\|\bm{Z}_k\|_{\mathcal{A}_k}(1-\|(\mathcal{B}^*\bm{\lambda})_k\|_{\mathcal{A}_k}^{\mathsf{d}})\Big].
\end{align}
Solving the latter optimization problem, we obtain 
\begin{align}
&L(\bm{\mathcal{Z}},\bm{\lambda})= \begin{cases}
\langle \bm{\lambda},\bm{y}\rangle,&~~ \|(\mathcal{B}^*\bm{\lambda})_k\|_{\mathcal{A}_k}^{\mathsf{d}}\le 1,~ k \in [K] \\
-\infty, &{\rm otherwise.}
\end{cases}	
\end{align}
By transforming implicit constraints into explicit ones, the dual problem becomes
\begin{align}\label{eq.lagran2}
\max_{\bm{\lambda}\in\mathbb{C}^{N}}~\langle \bm{\lambda},\bm{y}\rangle \quad {\rm s.t.}\quad 
\|(\mathcal{B}^*\bm{\lambda})_k\|_{\mathcal{A}_k}^{\mathsf{d}}\le 1,\quad 
k \in [K],
\end{align}
 where $\mathcal{B}^*:\mathbb{C}^{M}\rightarrow \bigoplus_{k=1}^K\mathbb{C}^{M_k\times N}$ denotes the adjoint operator of $\mathcal{B}$ and $\mathcal{B}^*\bm{\lambda}:=((\mathcal{B}^*\bm{\lambda})_k)_{k=1}^K$ is a matrix tuple where the $k$-th matrix is given by $(\mathcal{B}^*\bm{\lambda})_k=\sum\nolimits_{n=1}^{N}\lambda_n\bm{c}^k_n\bm{e}_n^{\mathsf{T}}.$
Maximization in \eqref{eq.lagran2} can also be presented in SDP format as 
\begin{equation}\label{prob.sdp}
    \begin{aligned}
    &\bm{\lambda}^* = \underset{\substack{\bm{\lambda}\in\mathbb{C}^{N}, \bm{Q}\in \mathbb{C}^{N\times N}}}{\rm argmax}~~{\rm Re}\big\{\langle \bm{\lambda}, \bm{y}\rangle\big\}\\
    &~~~~{\rm s.t.}~~
    \begin{bmatrix}
        \bm{Q}&(\mathcal{B}^{*}(\bm{\lambda}))_k\\
        (\mathcal{B}^{*}(\bm{\lambda}))_k^{\mathsf{H}}&\bm{I}_T
    \end{bmatrix}\succeq \bm{0}, \quad k \in [K],\\
    & ~~~~~~\langle \mathscr{T}({\bm{e}_q}), \bm{Q}  \rangle=1_{q=0}, \quad q=-N+1,..., N-1,
    \end{aligned}
\end{equation}	
 where $\mathscr{T}$ shows the Toeplitz structure. 
Maximization in \eqref{prob.sdp} is a convex problem; therefore,  it can be efficiently solved using the CVX toolbox \cite{grant2014cvx}.  Let $\hat{\bm{\lambda}}$ be the solution to the dual problem in~\eqref{eq.lagran2},
then the spikes can be localized by the peaks of the following term $ \hat{\mathcal{T}}_k = \Big\{\tau \in [0,1) | \big\|  (\mathcal{B}^* \hat{\bm{\lambda}})_k \bm{a}(\tau) \big\|_2=1 \Big\}.$ For instance, an example of this channel estimation is depicted in Fig~\ref{fig:Daul} for a case with $K=2$. 
 
 To recover the message vector and the channel amplitudes corresponding to user $k$, we form  $\widehat{\bm{Z}}_k=\sum\nolimits_{\ell}\widehat{g}_{\ell}^k \widehat{\bm{x}}_k\bm{a}(\widehat{\tau}_{\ell}^k)^{\mathsf{T}}$. Let $\widehat{\bm{g}}^k:=[\widehat{g}_1^k, \ldots,\widehat{g}_{P_k}^k]^{\mathsf{T}}$. Then, the rank one matrix $\widehat{\bm{x}}_k$ can be estimated as $\widehat{\bm{x}}_k\widehat{\bm{g}}^{k\mathsf{T}}=\widehat{\bm{Z}}_k\bm{A}^{(k)\dagger}$ where $\bm{A}^{(k)}:=[\bm{a}(\widehat{\tau}_k^1), \ldots,\bm{a}(\widehat{\tau}_k^{P_k})]^\mathsf{T}$. By using the assumption $\|\widehat{\bm{x}}_k\|_2=1$ and taking singular value decomposition, we can find $|\widehat{\bm{x}}_k|$ and $|\widehat{\bm{g}}^k|$ for $k=1, \ldots, K$.

\section{Simulation Results}\label{sec.simulations}
This section evaluates the optimization performance in \eqref{eq.lagran2} for different channel delays and message lengths. Then numerical experiments are implemented using MATLAB CVX Toolbox \cite{grant2014cvx}.  The delays' locations are generated uniformly at random with the minimum separation $\Delta\geq \frac{1}{N}$ to be smaller than what we theoretically expected.  The basis of low  
dimensional tall matrix $\bm{C}_k\in \mathbb{C}^{ N \times M_k}$ is generated uniformly at random for $k\in [K]$ from normal distribution $\mathcal{N}(0,1)$. The messages $\bm{x}_k, k=1, \ldots, K$ are generated \textit{i.i.d} and uniformly at random  from the unit sphere. Note that if there is some sort of coordination on the values of transmitted messages (e.g., positiveness) between users and \ac{BS}, \ac{GB2D} can unambiguously recover the transmitted messages.

For the first case,  we set $K=4$ with $P_1=\cdots=P_4=3$ from $N=200$ samples, and the filter size $M_k=5$ for $k\in [4]$. Also, the sensing matrix $\bm{D}$ is set to be identity, i.e., $\bm{D}=\bm{I}_N$. Then, the results are depicted in Fig~\ref{Fig(a):Circ}. Fig~\ref{Fig(a):Circ} shows that  the \ac{GB2D} can distinguish two closely spaced delays. In Fig~\ref{Fig(b):Circ}, we repeat this experiment for $K=3$ with $M_1=3,M_2=2,M_3=1$ from $N=128$ samples.  
\begin{figure}[!t]
\centering
 \begin{tikzpicture} 
    \begin{axis}[
        xlabel = {$N$},
        ylabel = {MSE},
        label style={font=\footnotesize},
        width=0.48\textwidth,
        height=5cm,
        xmin=40, xmax=120,
        ymin=1e-6, ymax=.5,
        legend style={nodes={scale=0.65, transform shape}, at={(0.3,0.85)}}, 
        ticklabel style = {font=\footnotesize},
        ymajorgrids=true,
        xmajorgrids=true,
        grid style=dashed,
        grid=both,
        ymode = log,
        grid style={line width=.1pt, draw=gray!10},
        major grid style={line width=.2pt,draw=gray!30},
    ]
    \addplot[smooth,
             thin,
        color=chestnut,
        line width=0.9pt,
        ]
    table[x=N,y=x1]
    {Data/Sim3.dat};
\addplot[ smooth,
             thin,
        color=airforceblue,
        line width=0.9pt,
        ]
    table[x=N,y=x2]
    {Data/Sim3.dat};
    \legend{$\|\widehat{\bm{x}}_1 - \bm{x}_1\|_{2}$, $\|\widehat{\bm{x}}_2 - \bm{x}_2\|_{2}$};
    \end{axis}
\end{tikzpicture}
  \caption{ Message recovery performance versus the number of samples $N$. Here, both transmitted signals have a message of size $4$, i.e.,  $M_1 = M_2= 4$. Moreover, we consider the number of channel multipath components as $P_1 = 5$ and $P_2 = 5$. }
   \label{fig:Mas}
\end{figure}
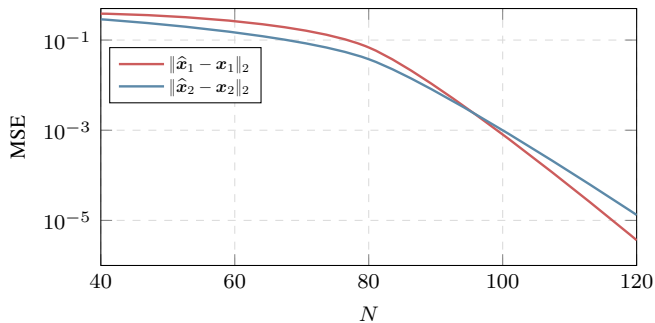


For the last case, we check the stability of the algorithm regarding a larger message size $M_k=16$ with $N=128$ samples and the effect of the sensing matrix with $M=64$ (sub-sampling uniform) in Fig~\ref{Fig(c):Circ}. We observe that all the cases are successfully recovered by \ac{GB2D} for a sufficient number of samples $N$. 

Finally, we evaluate the performance of the \ac{GB2D} method for message recovery in Fig~\ref{fig:Mas}.  Fig~\ref{fig:Mas} depicts the mean square error (MSE) of the estimated messages by the \ac{GB2D} for different numbers of samples $N$. We generate two messages of size $4$, i.e.,  $M_1 = M_2= 4$ with positive elements. The channel amplitudes are generated as complex Gaussian distribution, and the number of multipath components is set to $P_1 = 5$ and $P_2 = 5$. As it turns out from Fig. \ref{fig:Mas}, \ac{GB2D} provides excellent performance in message recovery, and the messages are unambiguously estimated\footnote{Note that the positiveness of the message elements and the $\ell_2$ normalized assumption of the message vector, i.e., $\|\bm{x}_k\|_2=1, k=1, \ldots, K$ are one of the ways to remove the multiplicative ambiguity caused by multiplying channel amplitudes and messages.}. Moreover, increasing the number of samples at the \ac{BS} leads to higher message recovery performance. 

The results show that for different numbers of multipath components and various sizes of messages, \ac{GB2D} can simultaneously recover messages and estimate multipath channels.

\section{Conclusion}\label{sec.conclusion}

The focus of our paper was to explore the possibility of simultaneous data recovery and channel estimation when multiple users send their messages through multiple channels and a \ac{BS} receives a linear combination of multiple convolved channels that are made up of a few scaled and delayed continuous Dirac spikes. Specifically, the measurements we worked with were a linear combination  of the collective sum of these convolved signals with unknown amplitudes.
The main goal was to find these unknown channel delay parameters from only one vector of observations. Since this problem is inherently highly challenging to solve, we overcome this issue by restricting the domain of the transmitted signals to some known low-dimensional subspaces while we have a separation condition on the Dirac spikes. Afterward, we proposed a semidefinite programming optimization to recover the channel delays and the messages of different users simultaneously from one observed vector.
For future works, we will provide the performance guarantee of \ac{GB2D} by obtaining the required sample complexity that one needs for perfect recovery of the continuous parameters and the transmitted waveforms.  Moreover, we plan to investigate further the impact of noise on the \ac{GB2D}'s performance and explore possible techniques to enhance its robustness.

\appendix

\section{Proof of Proposition \ref{prop.optimality}}\label{proof.optimality}

Any $\bm{\lambda}\in\mathbb{C}^N$ satisfying \eqref{eq.supp_cond} and \eqref{eq.offsupp_cond} is a feasible point in the dual problem \eqref{eq.lagran2}. Recall that $\bm{\mathcal{X}}=(\bm{X}_k)_{k=1}^K$ is the matrix tuple of interests where $\bm{X}_k=\sum_{\ell=1}^{P_k}g_{\ell}^k\bm{x}_k\bm{a}(\tau_{\ell}^k)^{\mathsf{T}}$. It holds that
\begin{align}\label{eq.rel1}
\sum_{k=1}^K\|\bm{X}_k\|_{\mathcal{A}_k}&\ge \sum_{k=1}^K\|\bm{X}_k\|_{\mathcal{A}_k}\|(\mathcal{B}^*\bm{\lambda})_k\|_{\mathcal{A}_k}^{\mathsf{d}}\nonumber\\ \nonumber
& \ge\sum_{k=1}^K\langle (\mathcal{B}^*\bm{\lambda})_k, \bm{X}_k \rangle\\ 
&=\sum_{k=1}^K\langle (\mathcal{B}^*\bm{\lambda})_k, \sum_{\ell=1}^{P_k}g_{\ell}^k\bm{x}_k\bm{a}(\tau_{\ell}^k)^{\mathsf{T}} \rangle\nonumber\\
&=\sum_{k=1}^K\sum_{\ell=1}^{P_k}{\rm Re}\Big\{{g_{\ell}^k}^*\langle (\mathcal{B}^{\ast}\bm{\lambda})_k, \bm{x}_k\bm{a}(\tau_{\ell}^k)^{\mathsf{T}} \rangle \Big\} \nonumber\\
&=\sum_{k=1}^K\sum_{\ell=1}^{P_k}{\rm Re}\Big\{{g_{\ell}^k}^*\langle \bm{Q}_k(\tau_{\ell}^k), \bm{x}_k \rangle\Big\},
\end{align}
where the second inequality is due to H\"{o}lder's inequality, and the last equalities are due to the definition of $\bm{Q}_k(\tau_{\ell}^k)$ in \eqref{eq.Qi}. We proceed \eqref{eq.rel1} by using conditions \eqref{eq.supp_cond} and \eqref{eq.offsupp_cond}:
\begin{align}\label{eq.rel2}
\sum_{k=1}^K\|\bm{X}_k\|_{\mathcal{A}_k}&\ge \sum_{k=1}^K\sum_{\ell=1}^{P_k}{\rm Re}\Big\{{g_{\ell}^k}^*\langle \frac{1}{\|\bm{x}_k\|_2^2}{\rm sgn}(g_{\ell}^k)\bm{x}_k, \bm{x}_k \rangle\Big\}  \nonumber\\
& = \sum_{k=1}^K\sum_{\ell=1}^{P_k}|g_{\ell}^k|\nonumber\\
&\ge\sum_{k=1}^K\|\bm{X}_k\|_{\mathcal{A}_k},
\end{align}
where we used the definition of atomic norm \eqref{eq.atomic_def} in the last step. From \eqref{eq.rel1} and \eqref{eq.rel2}, we find that 
$\langle \bm{\lambda}, \mathcal{B}\bm{\mathcal{X}}\rangle=\sum_{k=1}^K\|\bm{X}_k\|_{\mathcal{A}_k}$. Since the pair $(\bm{\mathcal{X}}, \bm{\lambda})$ is primal-dual feasible, we reach the conclusion that $\bm{\mathcal{X}}$ is an optimal solution of \eqref{eq.primalprob} and $\bm{\lambda}$ is an optimal solution of \eqref{eq.lagran2} by strong duality. For proving uniqueness, suppose $\widehat{\bm{\mathcal{X}}}:=(\widehat{\bm{X}}_k)_{k=1}^K$ is another optimal solution of \eqref{eq.primalprob} where $\widehat{\bm{X}}_k=\sum_{\widehat{\tau}_{\ell}^k\in \widehat{\mathcal{P}}_k}\widehat{g}_{\ell}^k\widehat{\bm{x}}_k\bm{a}(\widehat{\tau}_{\ell}^k)^{\mathsf{T}}$. If $\widehat{\bm{\mathcal{X}}}$ and $\bm{\mathcal{X}}$ have the same set of delays, i.e., $\widehat{\mathcal{P}}_k=\mathcal{P}_k,~\forall~k \in [K]$, we then have $\widehat{\bm{\mathcal{X}}}=\bm{\mathcal{X}}$ since the set of atoms building $\bm{\mathcal{X}}$ are linearly independent. If there exists some $\widehat{\tau}_{\ell}^k\notin \mathcal{P}_k$, then we can expand term $\langle \bm{\lambda}, \mathcal{B}\widehat{\bm{\mathcal{X}}}\rangle$ as follows 
\begin{align}\label{eq.rel3}
&\sum_{k=1}^K\langle (\mathcal{B}^*\bm{\lambda})_k, \widehat{\bm{X}}_k\rangle=\hspace{-4pt}
\sum_{k=1}^K \sum_{k}{\rm Re}\Big\{{\overset{*}{\widehat{g}_{\ell}^k}}\langle (\mathcal{B}^*\bm{\lambda})_k, \widehat{\bm{x}}_k\bm{a}(\widehat{\tau}_{\ell}^k)^{\mathsf{T}} \rangle\Big\}\hspace{-2pt}=\nonumber\\
&	{\footnotesize \sum_{k=1}^K\Big[\hspace{-6pt}\sum_{\widehat{\tau}_{\ell}^k\in \mathcal{P}_k}\hspace{-6pt}{\rm Re}\Big\{{\overset{*}{\widehat{g}_{\ell}^k}}
\langle\bm{Q}_k(\widehat{\tau}_{\ell}^k),\widehat{\bm{x}}_k \rangle\Big\}\hspace{-3pt}+\hspace{-10pt}\sum_{\widehat{\tau}_{\ell}^k\notin \mathcal{P}_k}\hspace{-6pt}{\rm Re}\Big\{{\overset{*}{\widehat{g}_{\ell}^k}}
\langle\bm{Q}_k(\widehat{\tau}_{\ell}^k),\widehat{\bm{x}}_k \rangle\Big\}\Big]\le}\nonumber\\
&\sum_{k=1}^K\Big[\sum_{\widehat{\tau}_{\ell}^k\in \mathcal{P}_k}\hspace{-6pt}|\widehat{g}_{\ell}^k|
\|\bm{Q}_k(\widehat{\tau}_{\ell}^k)\|_2\|\widehat{\bm{x}}_r\|_2 +\hspace{-7pt}\sum_{\widehat{\tau}_{\ell}^k\notin \mathcal{P}_k}\hspace{-5pt}|\widehat{g}_{\ell}^k|
\|\bm{Q}_k(\widehat{\tau}_{\ell}^k)\|_2\|\widehat{\bm{x}}_k\|_2\Big]\nonumber\\ \nonumber
&<\sum_{k=1}^K\Big[\sum_{\widehat{\tau}_{\ell}^k\in \mathcal{P}_k}|\widehat{g}_{\ell}^k|
 +\sum_{\widehat{\tau}_{\ell}^k\notin \mathcal{P}_k}|\widehat{g}_{\ell}^k|\Big]\\
 & =\sum_{k=1}^K|\widehat{g}_{\ell}^k|=\sum_{k=1}^K\|\widehat{\bm{X}}_k\|_{\mathcal{A}_k},
\end{align}
where we used the conditions \eqref{eq.supp_cond} and \eqref{eq.offsupp_cond}. The relation \eqref{eq.rel3} contradicts strong duality; hence $\bm{\mathcal{X}}$ is the unique optimal solution of \eqref{eq.primalprob}.

\bibliographystyle{IEEEtran}
\bibliography{Ref}

\begin{thebibliography}{10}
\providecommand{\url}[1]{#1}
\csname url@samestyle\endcsname
\providecommand{\newblock}{\relax}
\providecommand{\bibinfo}[2]{#2}
\providecommand{\BIBentrySTDinterwordspacing}{\spaceskip=0pt\relax}
\providecommand{\BIBentryALTinterwordstretchfactor}{4}
\providecommand{\BIBentryALTinterwordspacing}{\spaceskip=\fontdimen2\font plus
\BIBentryALTinterwordstretchfactor\fontdimen3\font minus
  \fontdimen4\font\relax}
\providecommand{\BIBforeignlanguage}[2]{{%
\expandafter\ifx\csname l@#1\endcsname\relax
\typeout{** WARNING: IEEEtran.bst: No hyphenation pattern has been}%
\typeout{** loaded for the language `#1'. Using the pattern for}%
\typeout{** the default language instead.}%
\else
\language=\csname l@#1\endcsname
\fi
#2}}
\providecommand{\BIBdecl}{\relax}
\BIBdecl

\bibitem{fodor2023optimizing}
S.~Fodor, G.~Fodor, D.~G{\"u}rg{\"u}no{\u{g}}lu, and M.~Telek, ``Optimizing
  pilot spacing in {MU-MIMO} systems operating over aging channels,''
  \emph{IEEE Trans. on Commun.}, 2023.

\bibitem{sayeed2010wireless}
A.~M. Sayeed and T.~Sivanadyan, ``Wireless communication and sensing in
  multipath environments using multiantenna transceivers,'' \emph{Handbook on
  Array Processing and Sensor Networks}, pp. 115--170, 2010.

\bibitem{SaeedBlind2022}
S.~Razavikia, J.~A. Peris, J.~M.~B. Da~Silva, and C.~Fischione, ``Blind
  asynchronous over-the-air federated edge learning,'' in \emph{IEEE Globecom},
  2022.

\bibitem{razavikia2023computing}
S.~Razavikia, J.~M. B.~d. Silva~Jr, and C.~Fischione, ``Computing functions
  over-the-air using digital modulations,'' in \emph{ICC}, 2023.

\bibitem{three-dimensionalsuper}
B.~Huang, W.~Wang, M.~Bates, and X.~Zhuang, ``Three-dimensional
  super-resolution imaging by stochastic optical reconstruction microscopy,''
  \emph{Science}, 2008.

\bibitem{SaeedBinary2020}
S.~Razavikia, A.~Amini, and S.~Daei, ``Reconstruction of binary shapes from
  blurred images via {Hankel}-structured low-rank matrix recovery,'' \emph{IEEE
  Trans. on Image Proc.}, 2020.

\bibitem{razavikia2019sampling}
S.~Razavikia, H.~Zamani, and A.~Amini, ``Sampling and recovery of binary shapes
  via low-rank structures,'' in \emph{2019 13th International conference on
  Sampling Theory and Applications (SampTA)}.\hskip 1em plus 0.5em minus
  0.4em\relax IEEE, 2019, pp. 1--4.

\bibitem{maskan2023demixing}
H.~Maskan, S.~Daei, and M.~H. Kahaei, ``Demixing sines and spikes using
  multiple measurement vectors,'' \emph{Signal Processing}, vol. 203, p.
  108786, 2023.

\bibitem{valiulahi2019two}
I.~Valiulahi, S.~Daei, F.~Haddadi, and F.~Parvaresh, ``Two-dimensional
  super-resolution via convex relaxation,'' \emph{IEEE Transactions on Signal
  Processing}, vol.~67, no.~13, pp. 3372--3382, 2019.

\bibitem{jung2017blind}
P.~Jung, F.~Krahmer, and D.~St{\"o}ger, ``Blind demixing and deconvolution at
  near-optimal rate,'' \emph{IEEE Trans. on Info. Theo.}, 2017.

\bibitem{daei2023blind}
S.~Daei and M.~Kountouris, ``Blind goal-oriented massive access for future
  wireless networks,'' \emph{IEEE Trans. on Sig. Proc.}, 2023.

\bibitem{vetterli2010multichannel}
M.~Vetterli and Y.~Lu, ``Multichannel sampling with unknown gains and offsets:
  A fast reconstruction algorithm,'' in \emph{Proceedings of Allerton
  Conference on Communication, Control and Computing}, 2010.

\bibitem{SayyariBlind2021}
S.~Sayyari, S.~Daei, and F.~Haddadi, ``Blind two-dimensional super-resolution
  in multiple-input single-output linear systems,'' \emph{IEEE Sig. Proc.
  Letters}, 2021.

\bibitem{daei2023blinda}
S.~Daei, S.~Razavikia, M.~Kountouris, M.~Skoglund, G.~Fodor, and C.~Fischione,
  ``Blind asynchronous goal-oriented detection for massive connectivity,'' in
  \emph{WiOPt}, 2023.

\bibitem{liu2020joint}
F.~Liu, C.~Masouros, A.~P. Petropulu, H.~Griffiths, and L.~Hanzo, ``Joint radar
  and communication design: Applications, state-of-the-art, and the road
  ahead,'' \emph{IEEE Trans. on Commun.}, 2020.

\bibitem{seidi2022novel}
M.~Seidi, S.~Razavikia, S.~Daei, and J.~Oberhammer, ``A novel demixing
  algorithm for joint target detection and impulsive noise suppression,''
  \emph{IEEE Commun. Letters}, vol.~26, no.~11, pp. 2750--2754, 2022.

\bibitem{safari2021off}
M.~Safari, S.~Daei, and F.~Haddadi, ``Off-the-grid recovery of time and
  frequency shifts with multiple measurement vectors,'' \emph{Signal
  Processing}, vol. 183, p. 108016, 2021.

\bibitem{jain2017non}
P.~Jain, P.~Kar \emph{et~al.}, ``Non-convex optimization for machine
  learning,'' \emph{Found. and Trends{\textregistered} in ML}, 2017.

\bibitem{ahmed2013blind}
A.~Ahmed, B.~Recht, and J.~Romberg, ``Blind deconvolution using convex
  programming,'' \emph{IEEE Trans. on Info. Theo.}, 2013.

\bibitem{ling2015self}
S.~Ling and T.~Strohmer, ``Self-calibration and biconvex compressive sensing,''
  \emph{Inverse Problems}, 2015.

\bibitem{candes2013phaselift}
E.~J. Candes, T.~Strohmer, and V.~Voroninski, ``Phaselift: Exact and stable
  signal recovery from magnitude measurements via convex programming,''
  \emph{Commus. on Pure and App. Math.}, 2013.

\bibitem{daei2019distribution}
S.~Daei, F.~Haddadi, and A.~Amini, ``Distribution-aware block-sparse recovery
  via convex optimization,'' \emph{IEEE Signal Processing Letters}, vol.~26,
  no.~4, pp. 528--532, 2019.

\bibitem{chi2016guaranteed}
Y.~Chi, ``Guaranteed blind sparse spikes deconvolution via lifting and convex
  optimization,'' \emph{IEEE Journal of Sel. Top. in Sig. Proc.}, 2016.

\bibitem{Jacome2020DualBlind}
R.~Jacome, K.~V. Mishra, E.~Vargas, B.~M. Sadler, and H.~Arguello,
  ``Multi-dimensional dual-blind deconvolution approach toward joint
  radar-communications,'' in \emph{IEEE SPAWC}, 2022.

\bibitem{ke2020compressive}
M.~Ke, Z.~Gao, Y.~Wu, X.~Gao, and R.~Schober, ``Compressive sensing-based
  adaptive active user detection and channel estimation: Massive access meets
  massive {MIMO},'' \emph{IEEE Trans. on Sig. Proc.}, 2020.

\bibitem{chandrasekaran2012convex}
V.~Chandrasekaran, B.~Recht, P.~A. Parrilo, and A.~S. Willsky, ``The convex
  geometry of linear inverse problems,'' \emph{Found. of Comp. Math.}, 2012.

\bibitem{grant2014cvx}
M.~Grant and S.~Boyd, ``{CVX}: {MATLAB} software for disciplined convex
  programming, version 2.1,'' 2014.

\end{thebibliography}

\end{document}